%% file: rounded-hartley-cleaned.tex
\RequirePackage[l2tabu, orthodox]{nag}
\RequirePackage{snapshot}

\documentclass[9pt,twocolumn]{extarticle}

\makeatletter
\if@twocolumn
  \usepackage[dvips,letterpaper,top=0.5in, bottom=0.5in, left=0.75in, right=0.5in,includefoot,heightrounded]{geometry}
\else
  \usepackage[dvips,letterpaper,margin=1in,includefoot,heightrounded]{geometry}
\fi

\usepackage{srcltx}

\usepackage[utf8]{inputenc}
\usepackage{fouriernc}

\usepackage{amsmath}
\usepackage{amssymb,amsfonts}
\usepackage{euscript}

\usepackage{abstract}

\usepackage{epsfig}
\usepackage[usenames,dvipsnames,svgnames,x11names]{xcolor}
\usepackage{subfigure}

\usepackage{booktabs}

\usepackage{setspace}
\usepackage{flushend}

\usepackage{float}
\floatstyle{ruled}
\newfloat{Program}{thp}{lop}

\usepackage{cite}

\usepackage[short,12hr]{datetime}

\usepackage{hyperref}

\newtheorem{proposition}{Proposition}
\newtheorem{definition}{Definition}

\newtheorem{conjecture}{Conjecture}

\newcommand{\nnorm}{\operatorname{\bar{\mu}}}
\newcommand{\hartley}{\stackrel{\mathcal{H}}{\longleftrightarrow}}
\newcommand{\hq}{{\pmb{{\mathsf{H}}}}}
\newcommand{\cas}{\operatorname{cas}}

\def\QED{\mbox{$\square$}}
\def\proof{\noindent{\it Proof:~}}
\def\endproof{\hspace*{\fill}~\QED\par\endtrivlist\unskip}

\makeatletter

\newcommand{\printtitle}{%
\makeatletter
\if@twocolumn

\twocolumn[%
  \maketitle
  \begin{onecolabstract}
    \myabstract
  \end{onecolabstract}
  \begin{center}
    \small
    \textbf{Keywords}
    \\\medskip
    \mykeywords
  \end{center}
  \bigskip
]
\saythanks
\else
  \maketitle
  \begin{onecolabstract}
    \myabstract
  \end{onecolabstract}
  \begin{center}
    \small
    \textbf{Keywords}
    \\\medskip
    \mykeywords
  \end{center}
  \bigskip
  \onehalfspacing
\fi
\makeatother
}

\title{%
Rounded Hartley Transform:  A Quasi-involution\thanks{Manuscript originally published
in 2002 at the International Telecommunications Symposium.
Readers are encouraged to access
newer results
at
\url{https://arxiv.org/abs/2007.02232}.
}}

\author{%
R.~J.~Cintra
\thanks{%
R J Cintra is with the
Signal Processing Group,
Universidade Federal de Pernambuco,
Recife, Brazil.
E-mail: \url{rjdsc@de.ufpe.br}}
\quad
H. M. de Oliveira
\thanks{%
H. M. de Oliveira is with the
Signal Processing Group,
Universidade Federal de Pernambuco,
Recife, Brazil.
E-mail: \url{hmo@de.ufpe.br}}
\quad
C. O. Cintra%
\thanks{%
C. O. Cintra was with the Department
of Physics and Mathematics,
Rural Federal University of Pernambuco,
Recife, Brazil.}
}

\date{}

\newcommand{\myabstract}{%
A new multiplication-free transform derived from DHT is introduced:
the RHT.
Investigations on the properties of the RHT led us to the
concept of weak-inversion.
Using new constructs, we show that RHT is not
involutional like the DHT, but
exhibits quasi-involutional property, a new definition
derived from the periodicity of matrices.
Thus instead of using the actual inverse transform,
the RHT is viewed as an involutional transform,
allowing the use of direct (multiplication-free) to evaluate the inverse.
A fast algorithm to compute RHT is presented.
This algorithm show embedded properties.
We also extended RHT to the two-dimensional case.
This permitted us to perform a preliminary analysis on
the effects of RHT on images.
Despite of some SNR loss, RHT can be very interesting for applications involving
image monitoring associated to decision making, such as military applications
or medical imaging.
}

\newcommand{\mykeywords}{%
DCT Approximation,
Fast algorithms,
Image compression
}

\begin{document}

\printtitle

\section{Introduction}
{D}{iscrete}
transforms have a significant role in digital signal processing.
A relevant example is the discrete Hartley transform (DHT), which offers many advantages over the more popular discrete Fourier transform.
To cite major advantages,
(i) DHT is a real-valued transform (no complex arithmetic is needed),
(ii) it possesses same formula for forward and inverse transform,
(iii) it has a computational equivalence to DFT~\cite{Heideman},
(iv) DHT shows high symmetry, which is desirable from the implementation point-of-view, and
(v) it is mathematically elegant.
These characteristics have motivated a lot of research to promote the use of DHT instead of DFT.
Thus DHT has hit many applications such as spectral analysis, convolution computation, adaptive filters, interpolation, communication systems and medical imaging~\cite{Paik}.
A representative reference list with the literature about the Hartley transform is found in~\cite{Olejniczak}.

Another important area of signal processing concerns with the
minimal complexity methods.
The class of multiplication-free discrete transforms, such as Walsh/Hadamard transform, has attracted much interest, since those transforms provide low computational complexity.
The multiplication-free paradigm was adopted by Reed~\emph{et alli} in the definition of the arithmetic Fourier transform~\cite{Reed}.
Recently an algorithm of this kind was proposed: the arithmetic Hartley transform~\cite{Cintra:Interpolate}.
An interesting approach was done by Bhatnagar: using Ramanujan numbers, another multiplication-free transform was invented~\cite{Bhatnagar}.
Approximation procedures are also being taken in consideration.
In a recent paper~\cite{Dee}, Dee-Jeoti proposed the approximate fast Hartley transform, though multiplicative complexity is not null.

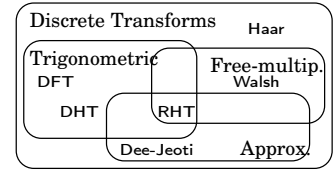
\begin{figure}
\centering
\input{qcas_e_as_transformadas.pstex_t}
\caption{Discrete transforms and some of their classes.
Rounded Hartley transform is placed in the intersection
of many classes.}
\label{rht-classes}
\end{figure}

Seeking for new procedures with the multiplication-free motto in mind, we introduced in this paper a new transformation: the rounded Hartley transform (RHT), a transform with zero multiplicative complexity.
Figure~\ref{rht-classes}  places
the RHT among other transforms.

In section~\ref{section-RHT} and~\ref{section-quasi-involution}, we define the RHT
and set the theoretical background to new concepts: quasi-equivalence, quasi-involution and weak-inversion, with simulation results.
Section~\ref{section-FRHT} brings a first approach to a fast algorithm for RHT, using the theoretical background found in~\cite{Oliveira:Factorization}.
We explored a naive example, 16-RHT, and derived arithmetic complexity bounds.
Subsequently in section~\ref{section-2DRHT}, the two-dimensional case was analyzed by introducing the 2-D RHT.
The effects of 2-D RHT on standard classical images were investigated, particularly peak signal-noise ratio.
We ended this paper establishing a connection between the new RHT and the Walsh/Hadamard transform.

\section{The Rounded Hartley Transform}
\label{section-RHT}

Let ${\mathbf{v}}$ be an $n$-dimensional vector with real elements.
The discrete Hartley transform establishes a pair of signal vectors
${\mathbf{v}} \hartley {\mathbf{V}}$, where the elements
of ${\mathbf{V}}$ are defined by
\begin{equation}
V_k
\triangleq
\sum_{i=0}^{n-1}
v_i
\cas
\left(
\frac{2\pi i k}{n}
\right),
\quad
k=0,1,\ldots,n-1,
\end{equation}
where $\cas(x) \triangleq \cos(x) + \sin(x)$.
This transform leads to the definition of Hartley
matrix~${\mathbf{H}}$, which elements are on the form
$h_{i,k}=\cas\left(\frac{2\pi i k}{n}\right)$.

The roundoff of a matrix
is obtained by rounding off
its elements.
Thus the rounded Hartley matrix elements
${\mathsf{h}}_{i,k}$ are defined by
\begin{equation}
{\mathsf{h}}_{i,k}
\triangleq
\left[
\underbrace{
\cas\left( \frac{2\pi  i k}{n}\right)
}_{h_{i,k}}
\right],
\quad
i,k=0,1,\ldots,N-1,
\end{equation}
where $[\cdot]$ denotes the nearest integer function.
For the sake of notation,
let us denote the rounded Hartley matrix by
$\hq$.

It is easy to see that
the elements ${\mathsf{h}}_{i,k}$ belong to $\{-1,0,1\}$,
since $|\cas(x)|\leq\sqrt2$.
Consequently, the rounded Hartley transform   can be implemented
using only additions, regardless the blocklength.
Rounded Hartley transform is a multiplication-free transform, which can be very attractive from the
practical point of view.

The first questions to be answered are:
(i) Is the spectrum derived from RHT a good estimation of the
(true) Hartley spectrum?
(ii) Is there an inverse Hartley transform?

\begin{figure}
\centering
\input{simple_example.latex}
\vspace{-.25cm}
\caption{A Simple example. Hartley spectrum evaluated by discrete Hartley transform~${\mathbf{V}}$
(filled line ---)
and rounded Hartley transform~${\mathsf{V}}$ (dotted line $\cdots$) of a vector~${\mathbf{v}}$
with 64 samples of function $f(x)=\cos(90\pi x)(x-\frac{1}{2})^2$ , $0\leq x\leq 1$.}
\label{simple:signal}
\end{figure}
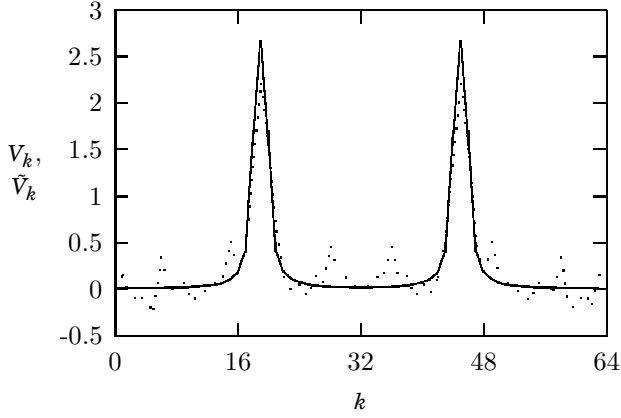

To begin with, we investigated the DHT and the RHT spectra for a few simple
signals (HT has real-valued components). %
Figure~\ref{simple:signal} shows both spectra for the signal
$f(x)=\cos(90\pi x)(x-\frac{1}{2})^2$ sampled by 64 points.
A pretty good agreement was observed.
A careful analysis of the error, or at least an upper bound, is currently being investigated.

\section {Involution and Quasi-involution}
\label{section-quasi-involution}

\begin{figure}
\centering
\subfigure[$n=2^4$]{ \epsfig{file=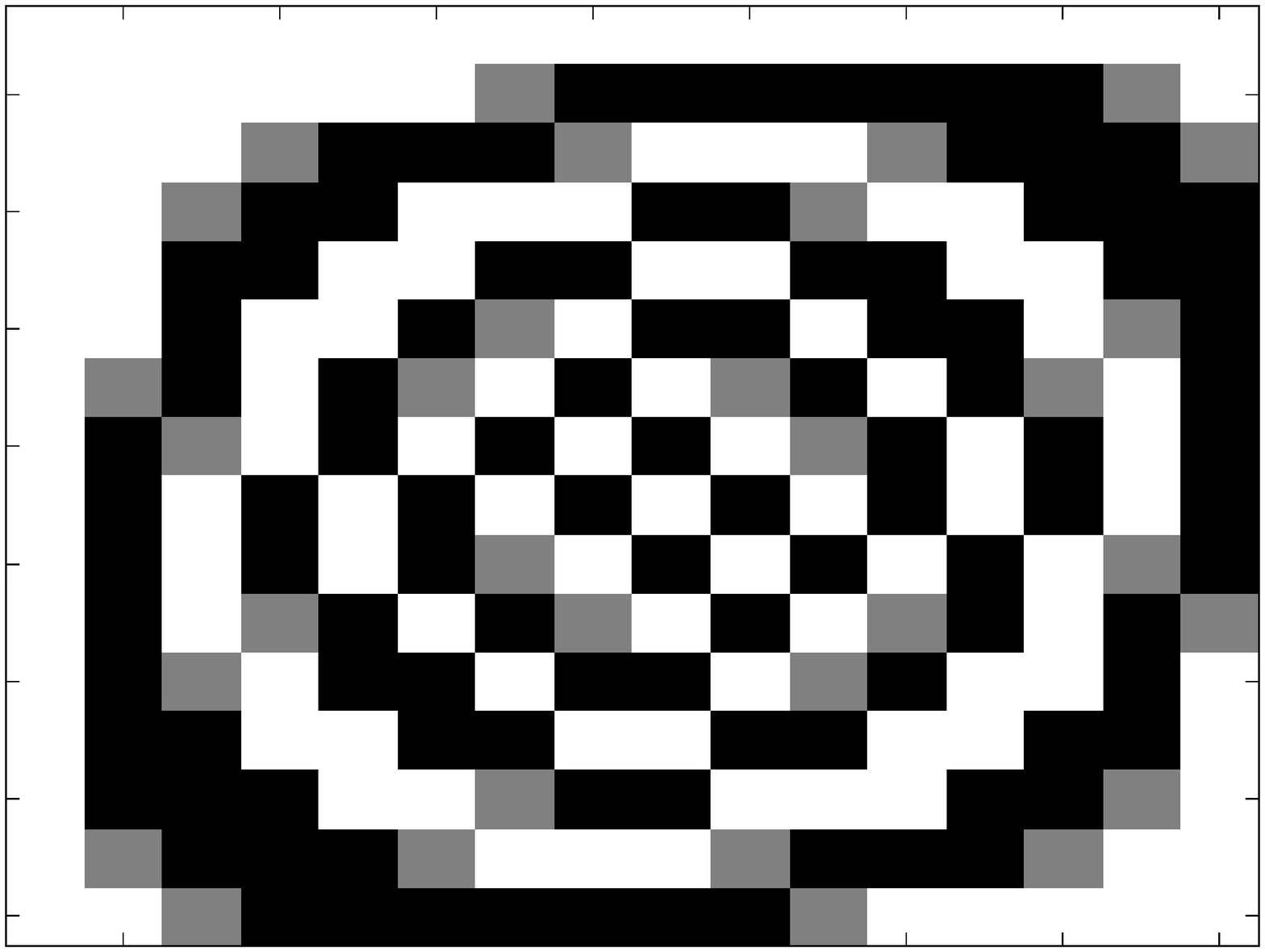,width=.22\linewidth,height=.22\linewidth}  }
\subfigure[$n=2^6$]{ \epsfig{file=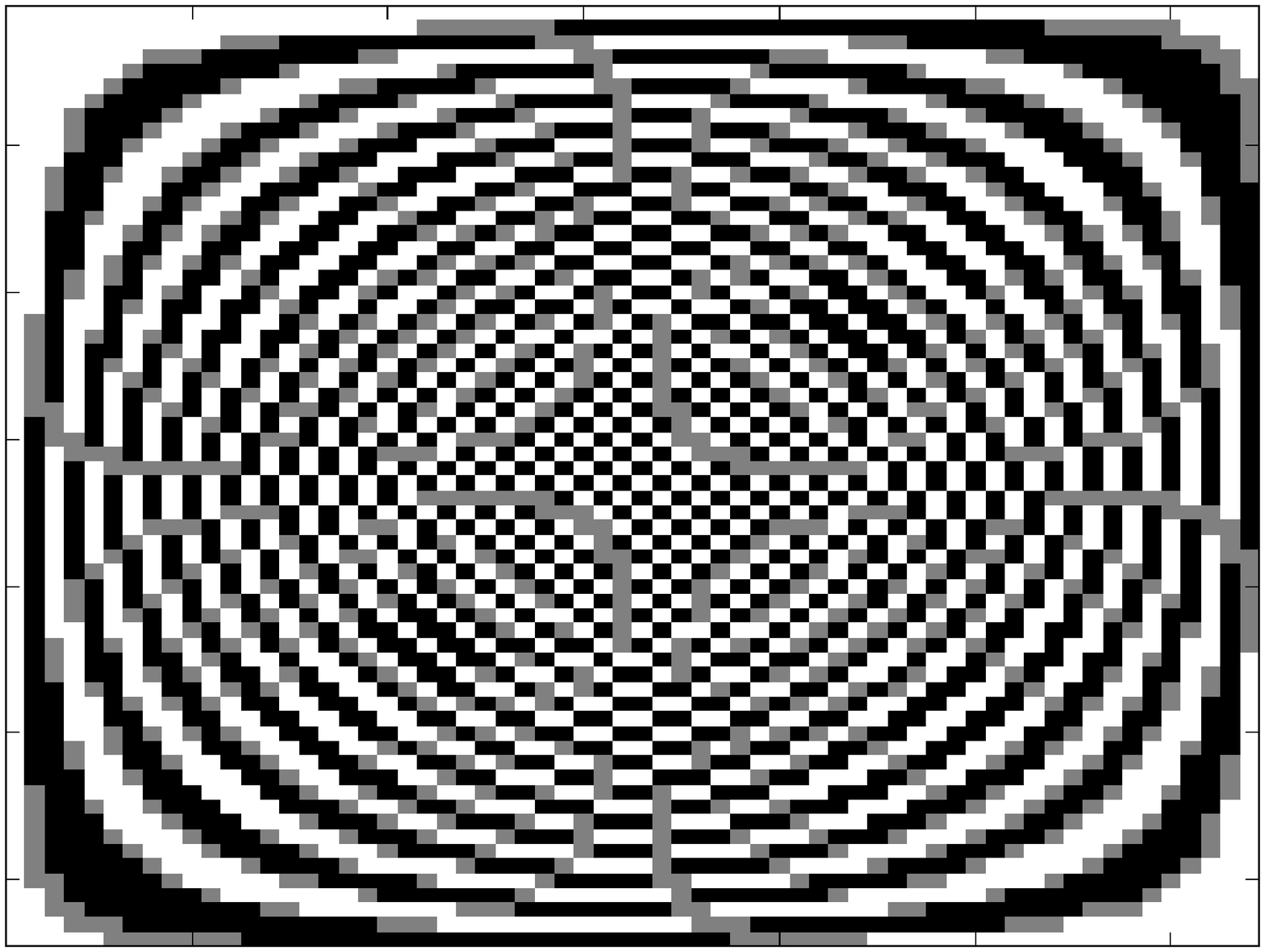,width=.22\linewidth,height=.22\linewidth}  }

\subfigure[$n=2^8$]{ \epsfig{file=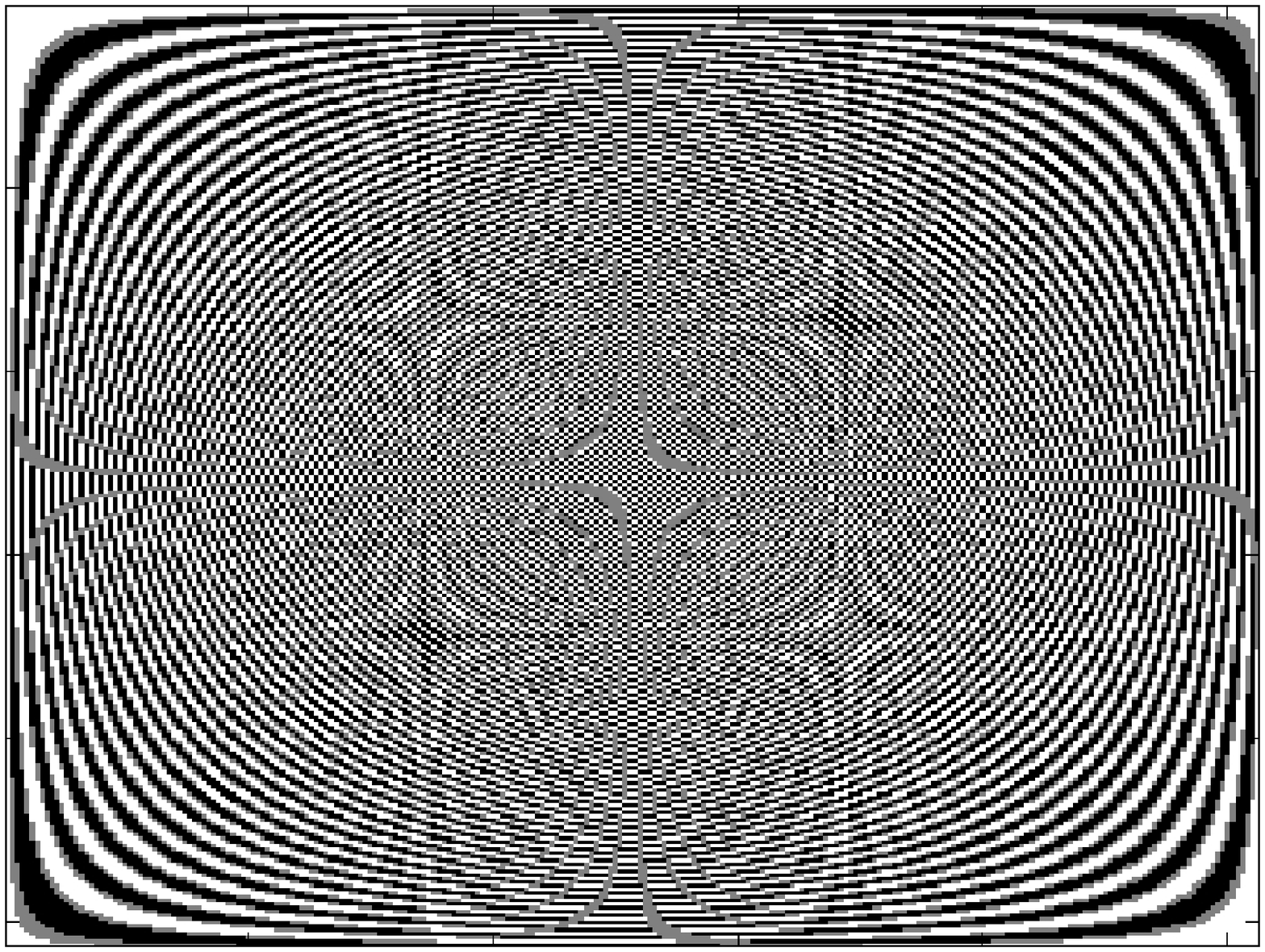,width=.22\linewidth,height=.22\linewidth}  }
\subfigure[$n=2^{10}$]{ \epsfig{file=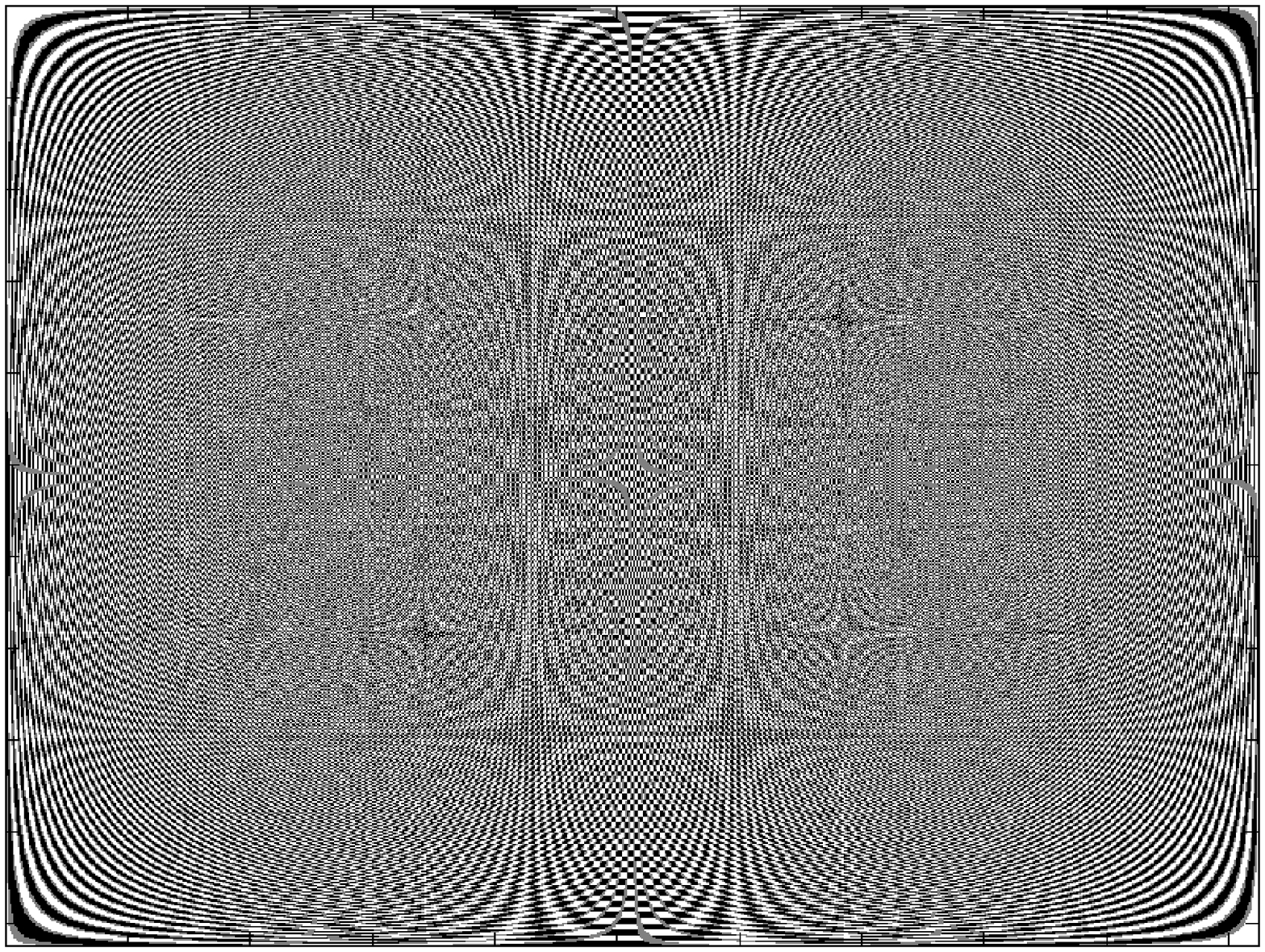,width=.22\linewidth,height=.22\linewidth}  }

\caption{Image patterns derived from rounded Hartley matrix for orders $n=16,64,256,1024$.
Each matrix $\hq$ is converted into intensity diagrams by representing
their elements in a gray scale.
Remark the presence of embedding patterns.}
\label{padrao:rht}
\end{figure}

In order to gain some insight on rounded Hartley transforms,
intensity diagrams were generated. The value ${\mathsf{h}}_{i,k}$
of each element of $\hq_n$ is converted into a gray-scale colormap
and the matrix $\hq_n$ is then represented by a square with $n^2$
pixels. %
Some interesting patterns derived from RHT are show in
Figure~\ref{padrao:rht}.

To keep rigorous with Hartley-Bracewell definition of the Hartley transform,
in this section, the Hartley matrix~${\mathbf{H}}_n$ and
the rounded Hartley transform~$\hq_n$ are scaled by $1/\sqrt{n}$.
Without any kind of conceptual loss,
this scaling does not interfere with the
results hereafter derived and
brings a greater elegance and harmony to the following constructs.

One of the most appealing properties of the classical DHT
is the fact that discrete Hartley matrix ${\mathbf{H}}_n$ is an involution,
i.e.,
${\mathbf{H}}_n^{-1} = {\mathbf{H}}_n$ (self-inverse).
However, after the round operation, $\cas(\cdot)$ kernel loses this characteristic
and
RHT is not an involution, since $\hq_n^{-1}\neq \hq_n$.

We found out by explicit computation
that the inverse of $\hq_n$ does exist for order
$n\leq 1024$.
Unfortunately $\hq_n^{-1}$ is not as interesting as $\hq_n$, since it
is computationally more intensive.
This fact was the key point that led us to a greater concern on
inverse matrices. We are particularly interested in finding out
matrices which have the properties of (i) being \emph{almost} the
inverse of a given matrix and (ii) being computationally more
interesting than the actual inverse.

That is, given a matrix~${\mathbf{A}}$ we are looking for a matrix
$\tilde{{\mathbf{A}}}$, such as:
\begin{equation}
{\mathbf{A}} \cdot \tilde{{\mathbf{A}}} \approx {\mathbf{I}}_n.
\end{equation}
This is what we called a \emph{weak-inverse}.

After further examination on  $\hq_n^{-1}$, we
observed that it resembles $\hq_n$ itself.
In fact, $\hq_n^{-1}$ is \emph{almost} $\hq_n$.
Since $\hq_n$ is defined from ${\mathbf{H}}_n$, we verify that $\hq_n$ is,
in some sense, \emph{almost} involutional %

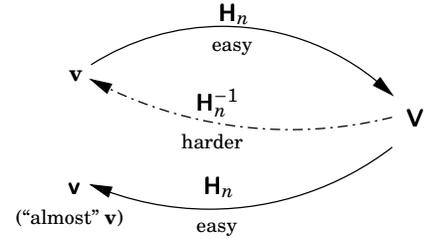
\begin{figure}
\centering
\input{diag_qcas.pstex_t}
\caption{Concept of weak-inversion. This diagram show the main idea behind our work.
We are concerned with matrices that ``almost'' invert. Filled arrows represent low
computational complexity, while the dashed one stands for a higher computational complexity.}
\label{diag:qcas}
\end{figure}

The qualitative idea was exposed, Figure~\ref{diag:qcas} may elucidate it.
Now it is time to imbue with some formalism this new concept
and establish strict definitions for the weak-inverse.

\subsection{Theory}

\begin{definition}
[Matrix Period]
The period of a matrix ${\mathbf{A}}$ is the
smallest positive integer $k$ such that
${\mathbf{A}}^{k+1}={\mathbf{A}}$.
\endproof
\end{definition}

For example, an idempotent transformation $T$ satisfies $T^2 = T$,
since it has period $k=1$.
An involution of a linear transformation
can be defined as a transformation which has period $k=2$.

\begin{definition}
The $n$-norm of a matrix  ${\mathbf{A}}$ is defined by
\begin{equation}
\nnorm ( {\mathbf{A}} ) = \frac{\| {\mathbf{A}}\|}{n},
\end{equation}
where $n$ is the order of ${\mathbf{A}}$ and
$\| \cdot \|$ represents Frobenius norm of a matrix:
$$
\|{\mathbf{A}}\| =
\left(
\sum_{i=1}^n
\sum_{j=1}^n
|a_{i,j}|^2
\right)^{1/2},
$$
where $a_{i,j}$ are the elements of matrix~${\mathbf{A}}$.
\endproof
\end{definition}

A family of matrices is defined as a sequence of matrices of increasing order,
which are generated by some rule.
A naive example is the identity family which is formed by all identity matrices:
$$
{\EuScript{I}}
=
\Big\{
{\mathbf{I}}_{1},
{\mathbf{I}}_{2},
{\mathbf{I}}_{3},
\ldots
\Big\}.
$$
Another example is the Fourier matrix family:
$$
{\EuScript{F}}
=
\Big\{
{\mathbf{F}}_1,
{\mathbf{F}}_2,
{\mathbf{F}}_3,
\ldots
\Big\}
$$
where matrices ${\mathbf{F}}_n$ have their elements defined according to
$$
f_{i,k} =
\exp
\left(
-j\frac{2\pi  i k}{n}
\right),
\quad
i,k=0\ldots,n-1.
$$
As a final example, the Hartley matrix family
$$
{\EuScript{H}}
=
\Big\{
{\mathbf{H}}_1,
{\mathbf{H}}_2,
{\mathbf{H}}_3,
\ldots
\Big\}
$$
is the sequence ${\mathbf{H}}_n$ of matrices
with elements defined by
$$
h_{i,k} =
\cas
\left(
\frac{2\pi i k}{n}
\right),
\quad
i,k=0\ldots,n-1.
$$

\begin{definition}
[Quasi-equivalence]
Two families of matrices  and ${\EuScript{B}}$
are said to be quasi-equivalent
at a level $\epsilon$
if and only if
for any two matrices ${\mathbf{A}}\in{\EuScript{A}}$ and ${\mathbf{B}}\in{\EuScript{B}}$
of same order
there exist a positive real number $\epsilon$ such as
\begin{equation}
\nnorm ({\mathbf{A}} - {\mathbf{B}} ) \leq \epsilon.
\end{equation}
\endproof
\end{definition}

Consider a family of matrices
${\EuScript{A}} = \Big\{ {\mathbf{A}}_1, {\mathbf{A}}_2, \ldots  \Big\}$.
Let us denote ${\EuScript{A}}_k$ the family which contains the
matrices of ${\EuScript{A}}$ raised to the $k$th power, i.e.,
${\EuScript{A}}_k = \Big\{ {\mathbf{A}}_1^k, {\mathbf{A}}_2^k, \ldots  \Big\}$.

\begin{definition}
[Quasi-periodicity]
\label{Quasi-periodicy}
The quasi-period of a matrix ${\mathbf{A}}_i\in{\EuScript{A}}$ is the smallest positive
integer $k$ such that
there exist a family of matrices
${\EuScript{A}}_k = \Big\{ {\mathbf{A}}_1^k, {\mathbf{A}}_2^k, \ldots, {\mathbf{A}}_i^k, \ldots  \Big\}$
which is
quasi-equivalent to
the identity family~${\EuScript{I}}$.

That is, for some $\epsilon$, the following equation is satisfied
\begin{equation}
\nnorm
\left(
{\mathbf{A}}^k_n - {\mathbf{I}}_n
\right)
\leq
\epsilon,
\qquad \forall n.
\end{equation}
\endproof
\end{definition}
The matrix ${\mathbf{A}}$ is said to be quasi-periodic with quasi-period~$k$.
A consequence of this definition is stated in the following proposition.

\begin{proposition}
All matrices in a family have the same quasi-period.
\end{proposition}
\proof
Let ${\EuScript{A}} = \left\{  {\mathbf{A}}_1,{\mathbf{A}}_2,{\mathbf{A}}_3, \ldots \right\}$
be a family of matrices.
Suppose that each matrix ${\mathbf{A}}_i$ has a given quasi-period $k_i$.
By Definition~\ref{Quasi-periodicy},
if a matrix~${\mathbf{A}}_n$ has quasi-period $k_n$, then there exists a
family
$$
{\EuScript{A}}_{k_n} =
\left\{
{\mathbf{A}}_1^{k_n},{\mathbf{A}}_2^{k_n},\ldots,{\mathbf{A}}_n^{k_n}, \ldots,{\mathbf{A}}_m^{k_n}, \ldots
\right\}
$$
which is quasi-equivalent to the identity family~${\EuScript{I}}$.
Now let another matrix~${\mathbf{A}}_m$ with quasi-period $k_m$.
Analogously, it implies that there exist a family
$$
{\EuScript{A}}_{k_m} =
\left\{
{\mathbf{A}}_1^{k_m},{\mathbf{A}}_2^{k_m},\ldots,{\mathbf{A}}_n^{k_m}, \ldots, {\mathbf{A}}_m^{k_m}, \ldots
\right\}
$$
which is quasi-equivalent to the identity family~${\EuScript{I}}$.

Since quasi-period is defined as the smallest integer for which the above conditions are
satisfied,
it yields that $k_n \leq k_m$ and $k_m \leq k_n$.
Thus $k_n = k_m$.
\endproof

After these theoretical background, we can formalize the concept of weak-inversion.
A matrix~$\tilde{{\mathbf{A}}}_n$ is a weak-inverse of~${\mathbf{A}}_n$
if the family of matrices
${\mathbf{A}}_n\tilde{{\mathbf{A}}}_n$
is
quasi-equivalent to the identity family, i.e.,
\begin{equation}
\label{weak-inverse}
\nnorm
\left(
{\mathbf{A}}_n
\tilde{{\mathbf{A}}}_n
-
{\mathbf{I}}_n
\right)
\leq
\epsilon
\quad
\forall
n.
\end{equation}

\begin{figure}
\centering
\input{erro_n.latex}
\vspace{-.25cm}
\caption{The $n$-norm of $\Big(\hq_n^2 - {\mathbf{I}}_n\Big)$
for $n=2,3,\ldots,1024$.}
\label{fig:erro_n}
\end{figure}

\subsection{The Weak-Inverse of Rounded Hartley Matrix}

Now let us get back to the rounded Hartley matrix~$\hq_n$.
Evaluating the $n$-norm of
$\hq^2_n - {\mathbf{I}}_n$
for $n=2,3,\ldots,1024$,
one can plot the graph depicted in Figure~\ref{fig:erro_n}.
After a data analysis of these points, we fitted them to
a Freundlich model:
\begin{equation}
\label{freudlich}
\nnorm
\left(
\hq^2_n - {\mathbf{I}}_n
\right)
\approx
a n^b,
\end{equation}
where
$a \approx 0.35167$ and $b \approx -0.49324$.

\begin{figure}
\centering
\subfigure[$n=56$]{ \fbox { \epsfig{file=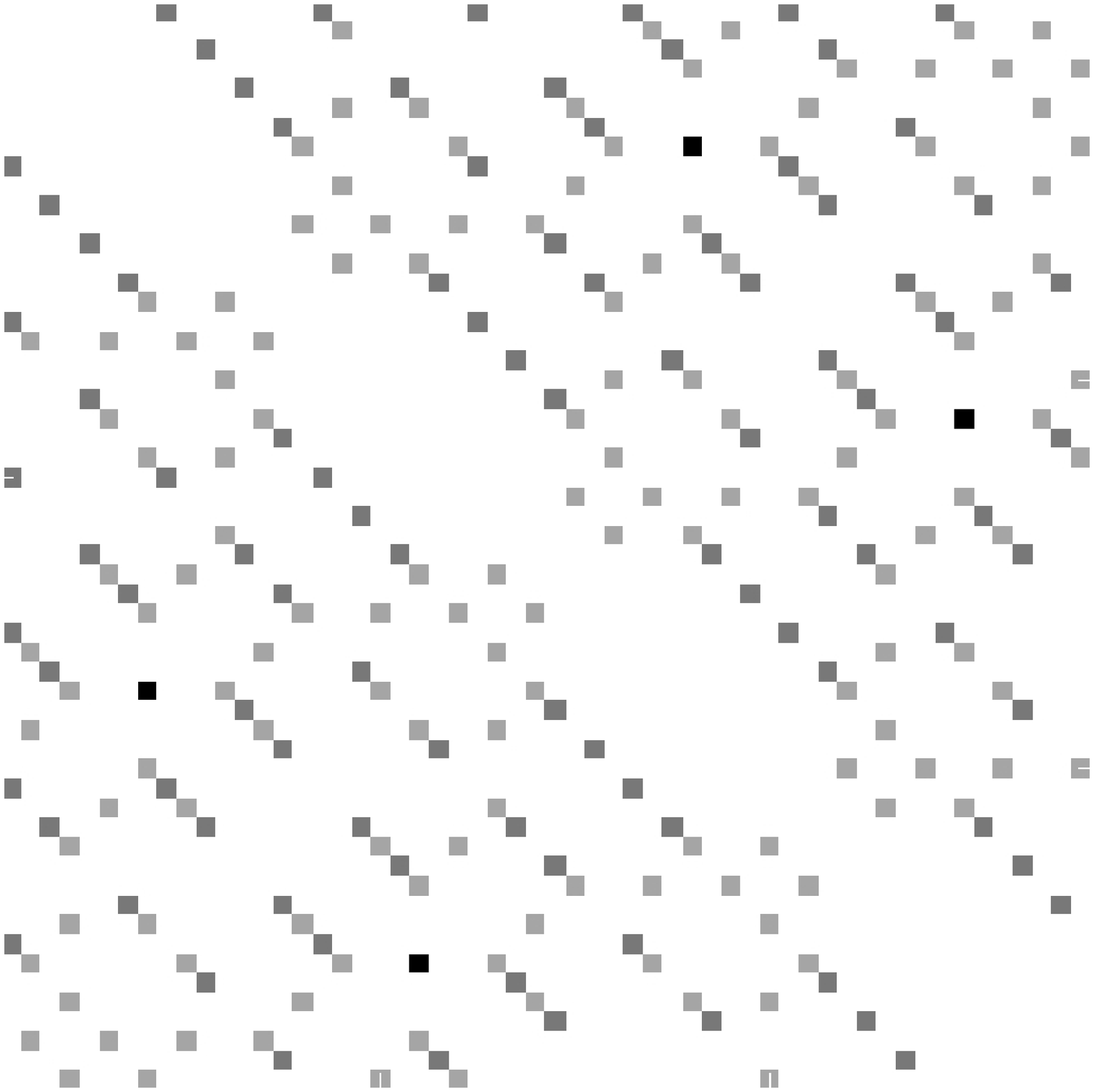,width=.41\linewidth,height=.41\linewidth} } }
\subfigure[$n=108$]{ \fbox { \epsfig{file=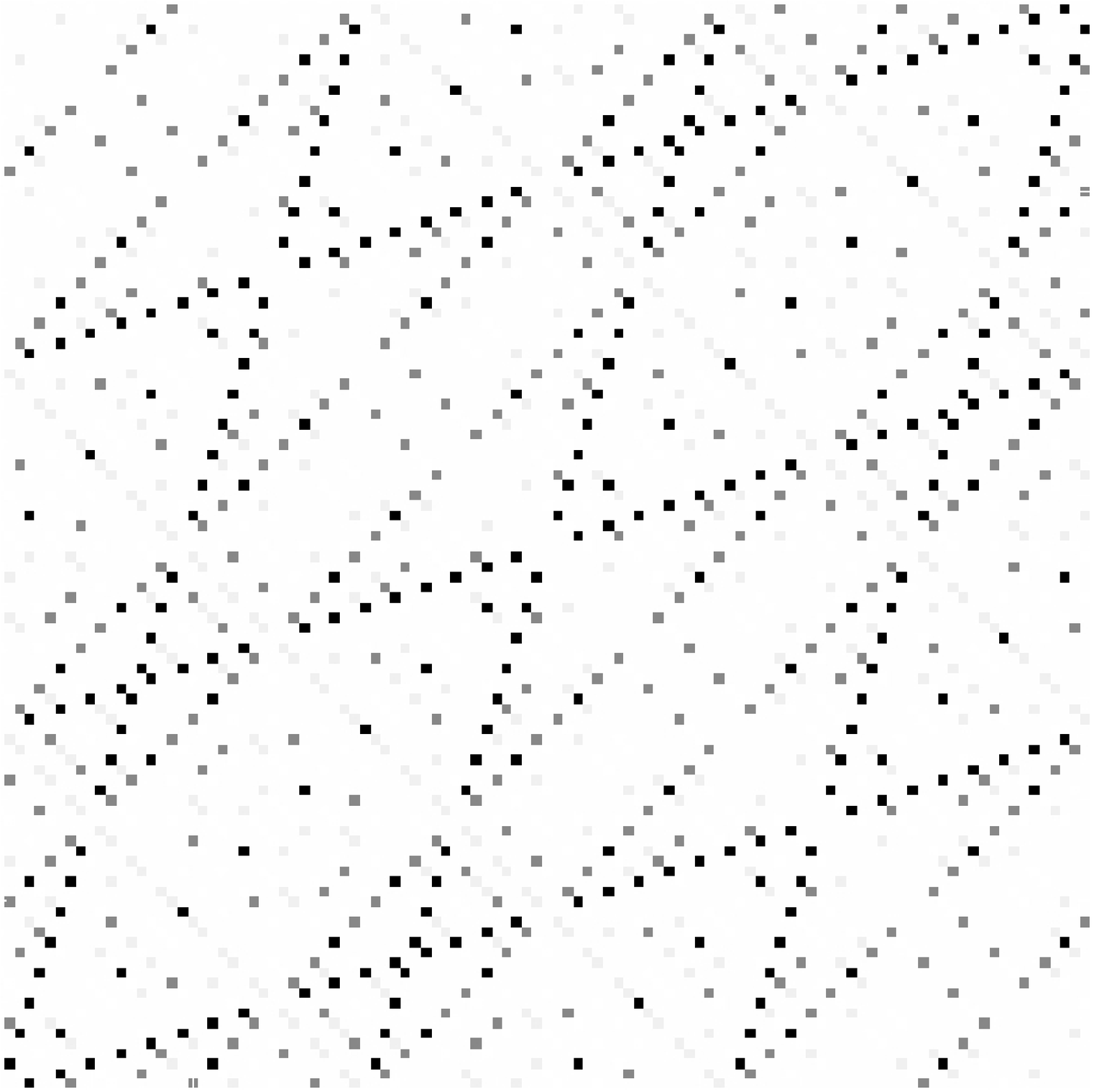,width=.41\linewidth,height=.41\linewidth} } }
\subfigure[$n=134$]{ \fbox{ \epsfig{file=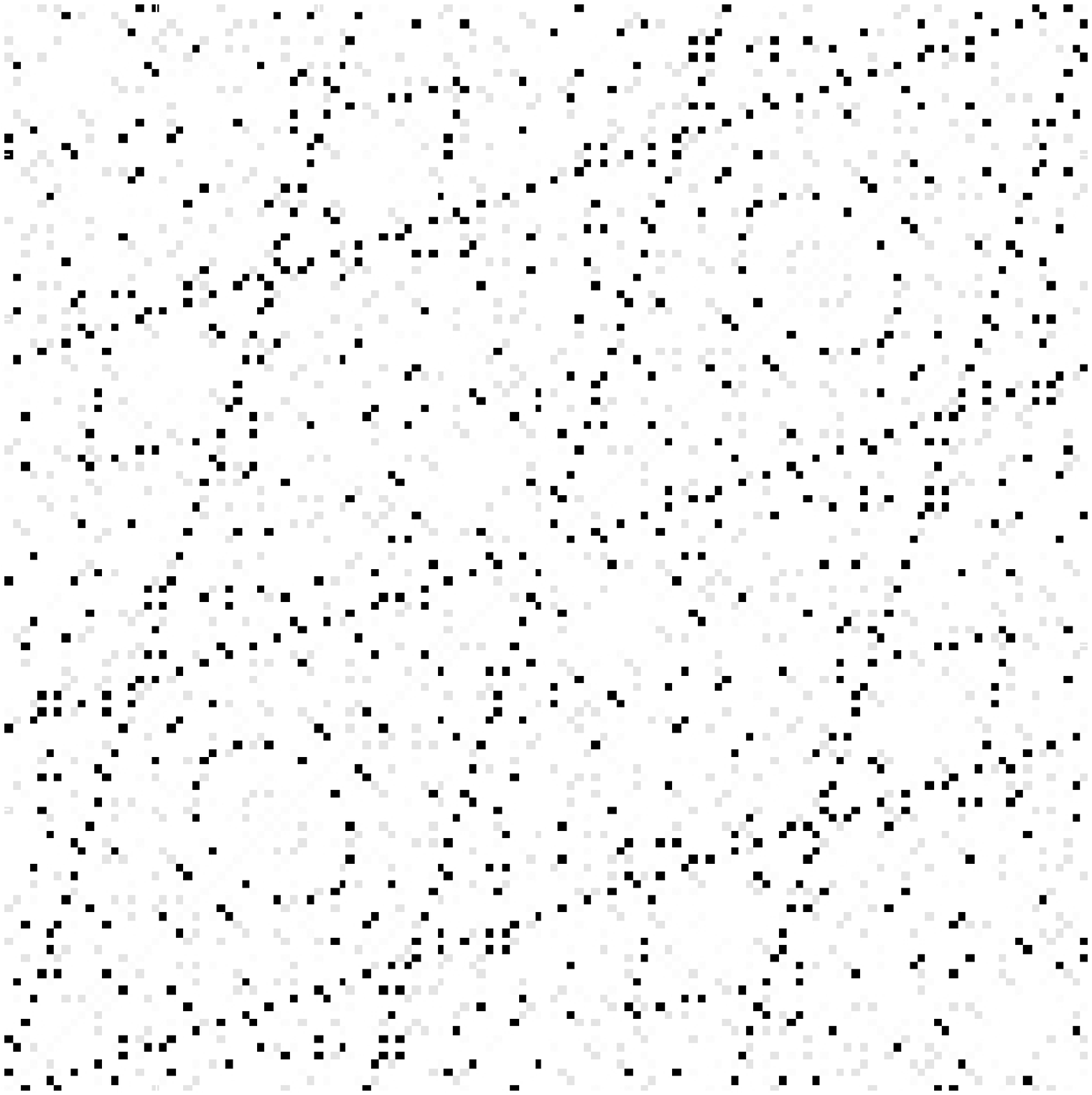,width=.41\linewidth,height=.41\linewidth} } }
\subfigure[$n=256$]{ \fbox{ \epsfig{file=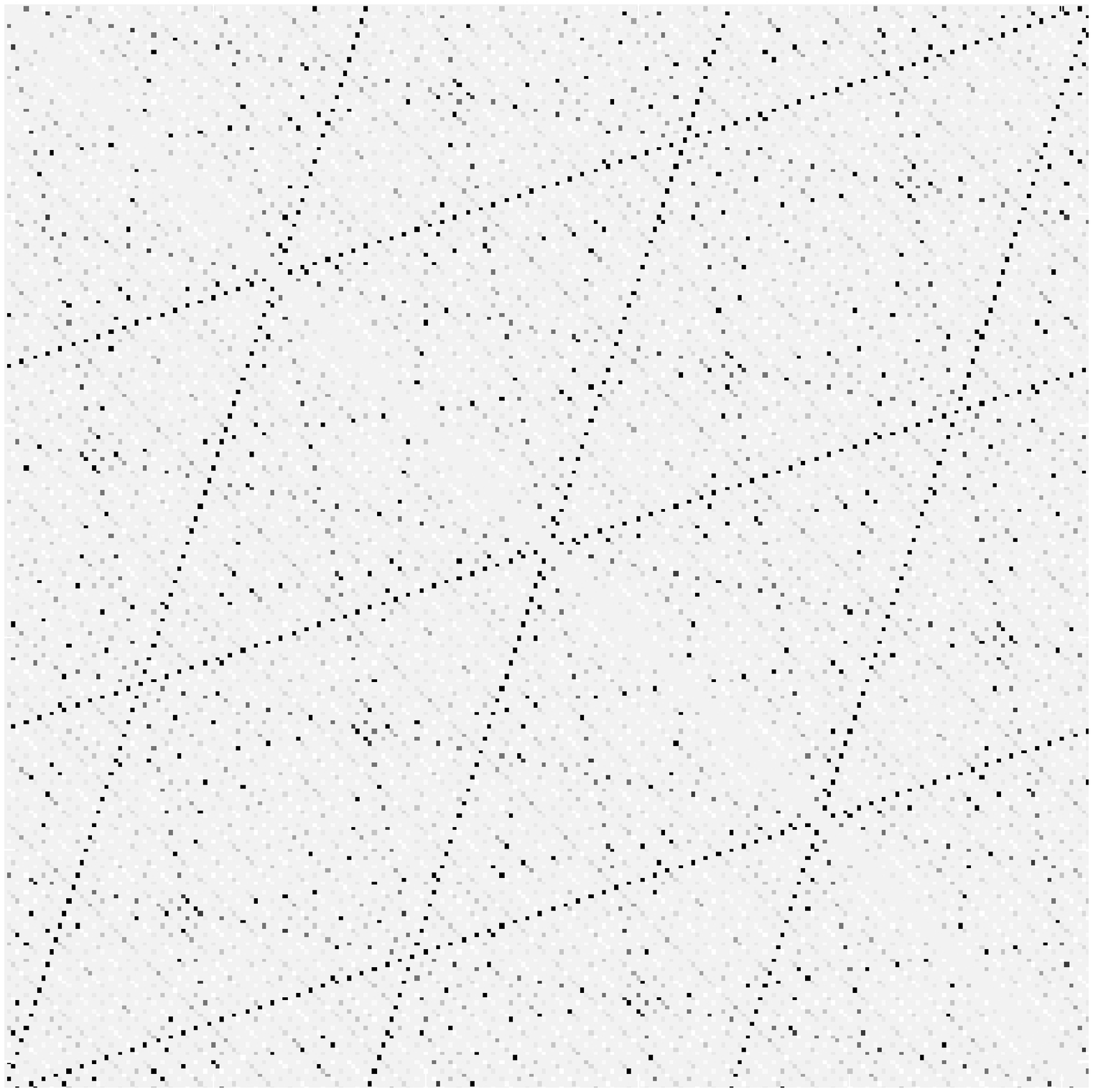,width=.41\linewidth,height=.41\linewidth} } }
\caption{Some interesting pictorial matrix patterns for $\hq^2_n$,  $n = 56$, $108$, $134$ and $256$.
A gray scale is used to plot the intensity of the elements:
the darker the element, the greater its magnitude (white denotes zeroes).
Main diagonal omitted for better visualization, since the magnitude of the
diagonal elements is much greater than the other elements'.
}
\label{fractal}
\end{figure}

These observations are the background needed to
infer on the asymptotical behavior
of $\nnorm\left( \hq^2_n - {\mathbf{I}}_n\right)$
and state the following conjecture:
\begin{conjecture}
\begin{equation}
\lim_{n\to\infty}
\nnorm
\left(
\hq^2_n - {\mathbf{I}}_n
\right)
=
0.
\end{equation}
\endproof
\end{conjecture}

\begin{sloppypar}
An immediate consequence of this conjecture is that
$\nnorm\left( \hq^2_n - {\mathbf{I}}_n\right)$ is bounded
and has its maximum value at $n=3$ (see Figure~\ref{fig:erro_n}).
That is,
$$
\nnorm
\left( \hq^2_n - {\mathbf{I}}_n\right)
\leq
\frac{2}{9},
\quad
n = 2, 3, \ldots
$$
According to Equation~\ref{weak-inverse},
it comes to uses that $\hq_n$ is
the weak-inverse of itself.
Furthermore under conditions
of Definition~\ref{Quasi-periodicy},
we can state that
 rounded Hartley matrices are
quasi-periodic and
their quasi-period is two.
\end{sloppypar}

\begin{definition}
[Quasi-involution]
A quasi-involution  is a transform with quasi-period of 2.
\endproof
\end{definition}

\subsection{General Comments}

In this subsection, we state some initial
observations
about the rounded Hartley transform
without further derivations or proofs.

\subsubsection{Error}
Since a weak-inverse is not precisely the inverse of a given matrix,
this approach of retrieving data from a weak-inverse introduces some degradation,
as expected.
The RHT is given by ${\pmb{\mathsf{V}}} = \hq_n {\mathbf{v}}$.
We shall use ${\pmb{\mathsf{v}}}=\hq_n {\pmb{\mathsf{V}}} = \hq_n^2 {\mathbf{v}}$
to compute the inverse, instead of the exact inverse RHT
${\mathbf{v}} = \hq_n^{-1}{\pmb{\mathsf{V}}}$.
Thus this procedure introduces an error by the use of the weak-inverse.
The error is therefore
\begin{equation}
{\pmb{\mathsf{v}}}-{\mathbf{v}}
=
\left(\hq^2_n - {\mathbf{I}}_n\right)
{\mathbf{v}}.
\end{equation}
As we see, the error ${\pmb{\mathsf{v}}}-{\mathbf{v}}$ depends on $\hq^2_n - {\mathbf{I}}_n$,
as well as on the original message~${\mathbf{v}}$.

\subsubsection{Fractal}
Since the measure $\nnorm\left(\hq^2_n - {\mathbf{I}}_n\right)$
presents a fractional exponent (Equation~\ref{freudlich}), objects
$\hq^2_n - {\mathbf{I}}_n$ should be
associated with some fractal.
The patterns displayed in Figure~\ref{fractal} show a
kind of self-similar behavior, as expected.

\section{A Fast Algorithm for RHT}
\label{section-FRHT}

In order to derive a fast algorithm, we use a naive example:
16-point RHT, which transform matrix~$\hq_{16}$ is shown below:
$$
{\hq}_{16}
=
\left[
\begin{smallmatrix}
 1 &  1 &  1 &  1 &  1 &  1 &  1 &  1 &  1 &  1 &  1 &  1 &  1 &  1 &  1 & 1 \\
 1 &  1 &  1 &  1 &  1 &  1 &    &  - &  - &  - &  - &  - &  - &  - &    & 1 \\
 1 &  1 &  1 &    &  - &  - &  - &    &  1 &  1 &  1 &    &  - &  - &  - &   \\
 1 &  1 &    &  - &  - &  1 &  1 &  1 &  - &  - &    &  1 &  1 &  - &  - & - \\
 1 &  1 &  - &  - &  1 &  1 &  - &  - &  1 &  1 &  - &  - &  1 &  1 &  - & - \\
 1 &  1 &  - &  1 &  1 &  - &    &  1 &  - &  - &  1 &  - &  - &  1 &    & - \\
 1 &    &  - &  1 &  - &    &  1 &  - &  1 &    &  - &  1 &  - &    &  1 & - \\
 1 &  - &    &  1 &  - &  1 &  - &  1 &  - &  1 &    &  - &  1 &  - &  1 & - \\
 1 &  - &  1 &  - &  1 &  - &  1 &  - &  1 &  - &  1 &  - &  1 &  - &  1 & - \\
 1 &  - &  1 &  - &  1 &  - &    &  1 &  - &  1 &  - &  1 &  - &  1 &    & - \\
 1 &  - &  1 &    &  - &  1 &  - &    &  1 &  - &  1 &    &  - &  1 &  - &   \\
 1 &  - &    &  1 &  - &  - &  1 &  - &  - &  1 &    &  - &  1 &  1 &  - & 1 \\
 1 &  - &  - &  1 &  1 &  - &  - &  1 &  1 &  - &  - &  1 &  1 &  - &  - & 1 \\
 1 &  - &  - &  - &  1 &  1 &    &  - &  - &  1 &  1 &  1 &  - &  - &    & 1 \\
 1 &    &  - &  - &  - &    &  1 &  1 &  1 &    &  - &  - &  - &    &  1 & 1 \\
 1 &  1 &    &  - &  - &  - &  - &  - &  - &  - &    &  1 &  1 &  1 &  1 & 1
\end{smallmatrix}
\right],
$$
where ``$-$'' represents $-1$ and blank spaces are zeroes.

Using methods described in~\cite{Oliveira:Factorization},
we obtained the implementation diagram displayed in Figure~\ref{fig:diagram}.

The algorithm turns out to have embedding properties: shorter transforms are found in major ones.
In the 16-point RHT, the following transforms are embedded: 2-, 4- and 8-point RHT.
By zeroing some inputs, a shorter transform is promptly available (e.g.
let $v_8 = \cdots = v_{15} = 0$ to have an 8-point RHT).
This feature makes the algorithm particularly flexible to
a much larger range of applicabilities~\cite{Oliveira:Factorization,Bracewell2}.

\begin{figure}

\centering
\input{qdht-16.pstex_t}
\caption{Flow graph for the fast algorithm of 16-point rounded Hartley transform.
Note the complete absence of multipliers.
The dashed boxes denote shorter transforms embedded in 16-point RHT.}
\label{fig:diagram}
\end{figure}

For blocklengths which are power of two, one can find out the the following arithmetic
complexity:
\begin{align}
A(n)&={\mathcal{O}}(n \log_2 n),\\
M(n)&=0,
\end{align}
where ${\mathcal{O}}(\cdot)$ is the Landau symbol.

\section{2-D Rounded Hartley Transform}
\label{section-2DRHT}

Original two-dimensional Hartley transform of an $n\times n$ image is defined by
\begin{equation}
b_{u,v}
=
\sum_{i=0}^{n-1}
\sum_{j=0}^{n-1}
a_{i,j}
\cdot
\cas
\left(
\frac{ui+vj}{n}
\right),
\end{equation}
where
$a_{i,j}$ are the elements of an image~${\mathbf{A}}$ and
$b_{u,v}$ are the elements of the Hartley transform of~${\mathbf{A}}$.

Since $\cas(\cdot)$ kernel is not separable, we cannot
express the two-dimensional transform in terms of a single
matrix equation, like the 2-D discrete Fourier transform.
Thus, we defined the
two-dimensional rounded Hartley transform
similarly to Bracewell's  method for two-dimensional
discrete Hartley transform~\cite{Bracewell:2d}.

Let ${\mathbf{A}}$ be the $n \times n$ image matrix.
We start the procedure by calculating
a temporary matrix~${\mathsf{T}}$, as follows:
\begin{equation}
{\pmb{\mathsf{T}}} = \hq_n \cdot {\mathbf{A}} \cdot \hq_n,
\end{equation}
where  $\hq_n$ is the rounded Hartley matrix of order $n$.
This is equivalent to take one-dimensional Hartley transform
of the rows, and then transform the columns~\cite{Gonzalez}.

Establishing that the elements of ${\pmb{\mathsf{T}}}$ are
represented on the form ${\mathsf{t}}_{i,j}$, $i,j=0,\ldots,n-1$,
we can  consider  three new matrices built from the
temporary matrix~${\pmb{\mathsf{T}}}$:
${\pmb{\mathsf{T}}}^{(c)}$,
${\pmb{\mathsf{T}}}^{(r)}$
and
${\pmb{\mathsf{T}}}^{(cr)}$ which elements are
${\mathsf{t}}_{i , n-j \pmod n}$,
${\mathsf{t}}_{n-i \pmod n, j}$,
${\mathsf{t}}_{n-i \pmod n, n-j \pmod n}$, respectively.
These different indexes flip matrix~${\pmb{\mathsf{T}}}$ in left-right direction,
except from the first column (${\pmb{\mathsf{T}}}^{(c)}$);
in up-down direction, except from the first line
(${\pmb{\mathsf{T}}}^{(r)}$);
and
both operations at same time (${\pmb{\mathsf{T}}}^{(cr)}$).

Using these constructs, the rounded Hartley transform~${\pmb{\mathsf{B}}}$
of an $n\times n$ image ${\mathbf{A}}$ is
defined as
\begin{equation}
{\pmb{\mathsf{B}}}
\triangleq
{\pmb{\mathsf{T}}}+
{\pmb{\mathsf{T}}}^{(c)}+
{\pmb{\mathsf{T}}}^{(l)}-
{\pmb{\mathsf{T}}}^{(cl)}.
\end{equation}

This definition derives directly from the $\cas(\cdot)$ property:
$
\cas(a+b) =
\cas(a)\cas(b)+
\cas(a)\cas(-b)+
\cas(-a)\cas(b)-
\cas(-a)\cas(-b)
$.

\begin{Program}
\caption{A simple \textsc{Matlab} program to compute 2-D RHT, its weak-inverse and the PSNR.}
\label{program:matlab}
{\scriptsize
\begin{verbatim}
function Z = rcas(N)

i = 0:(N-1);
j = 0:(N-1);
[I,J] = meshgrid(i,j);
Z = round ( cas ( 2 * pi / N * I .* J) );
\end{verbatim}
\hrulefill
\begin{verbatim}
function [B, AA, PSNR] = twodrht(file)

A = imread(file ,'bmp');
A = double(A);
[M, N] = size(A);
if M ~= N end;
colormap(gray(256));
K = rcas(N);
TEMP = K * A * K;
TEMPFLIPCOL = [TEMP(:,1),fliplr(TEMP(:,2:N))];
TEMPFLIPROW = [TEMP(1,:);flipud(TEMP(2:N,:))];
TEMPFLIPRC = [TEMPFLIPCOL(1,:);flipud(TEMPFLIPCOL(2:N,:))];
B = (1/2)*(TEMP+TEMPFLIPCOL+TEMPFLIPROW-TEMPFLIPRC);
temp = (1/N) * (1/N) * K * B * K;
tempFLIPCOL = [temp(:,1),fliplr(temp(:,2:N))];
tempFLIPROW = [temp(1,:);flipud(temp(2:N,:))];
tempFLIPRC = [tempFLIPCOL(1,:);flipud(tempFLIPCOL(2:N,:))];
AA  = (1/2)*(temp+tempFLIPCOL+tempFLIPROW-tempFLIPRC);
MSE = (1/N^2) * sum(sum((AA-A).^2));
RMSE = sqrt(MSE);
PSNR = 20 * log10(255/RMSE);
\end{verbatim}
}
\end{Program}

\begin{figure}
\centering
\subfigure[Moon surface]{ \epsfig{file=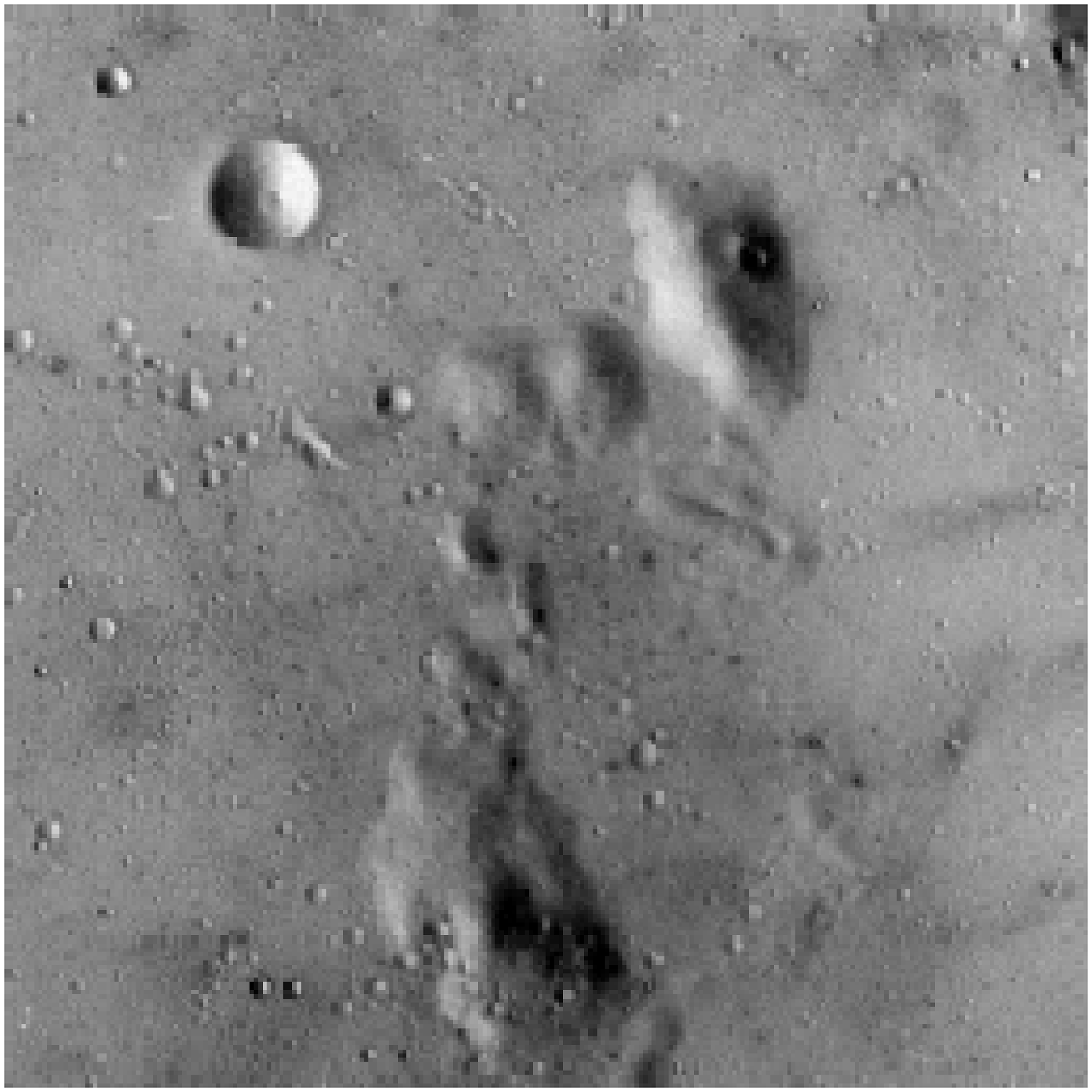,width=.45\linewidth}
\quad
\epsfig{file=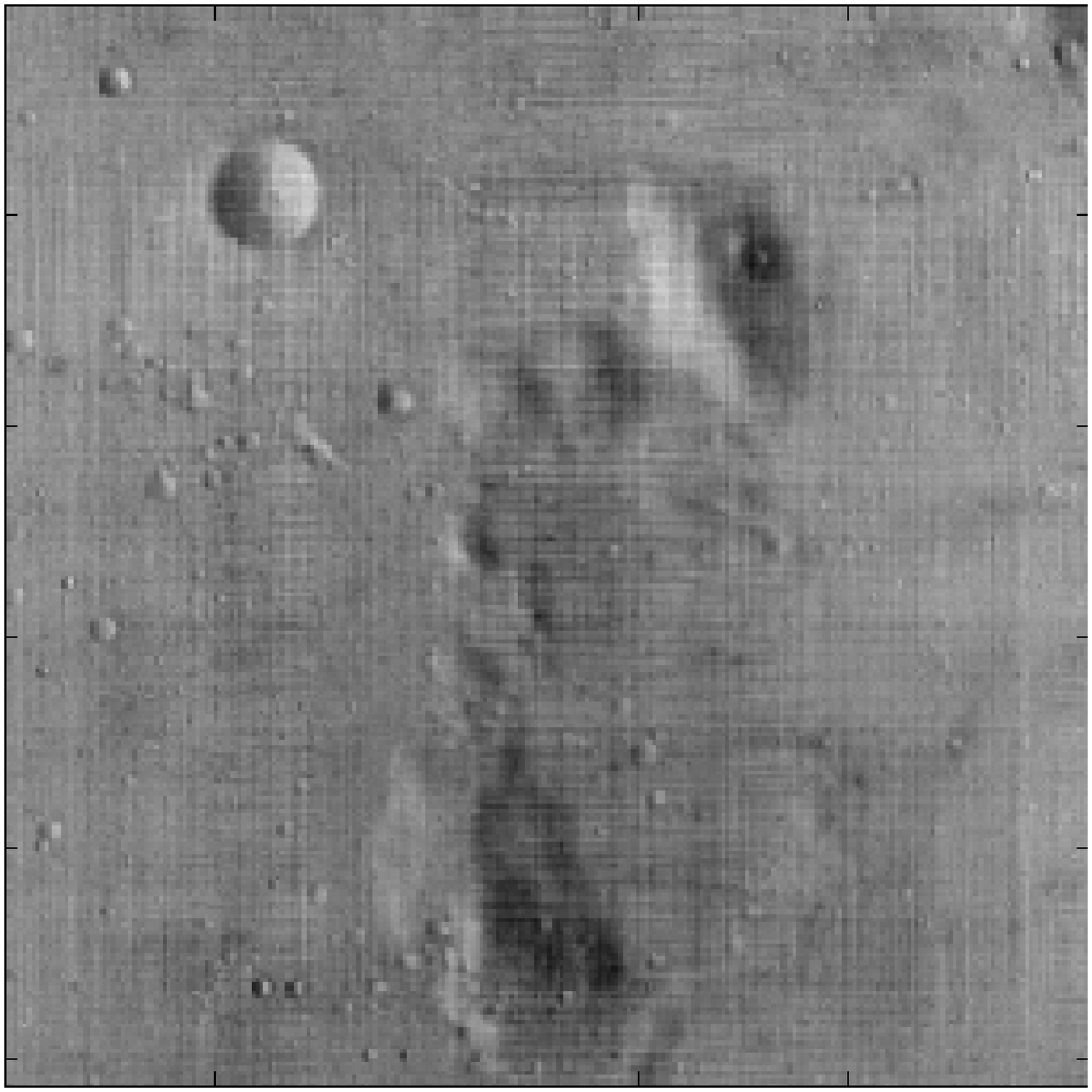,width=.45\linewidth}  }
\subfigure[Airplane]{ \epsfig{file=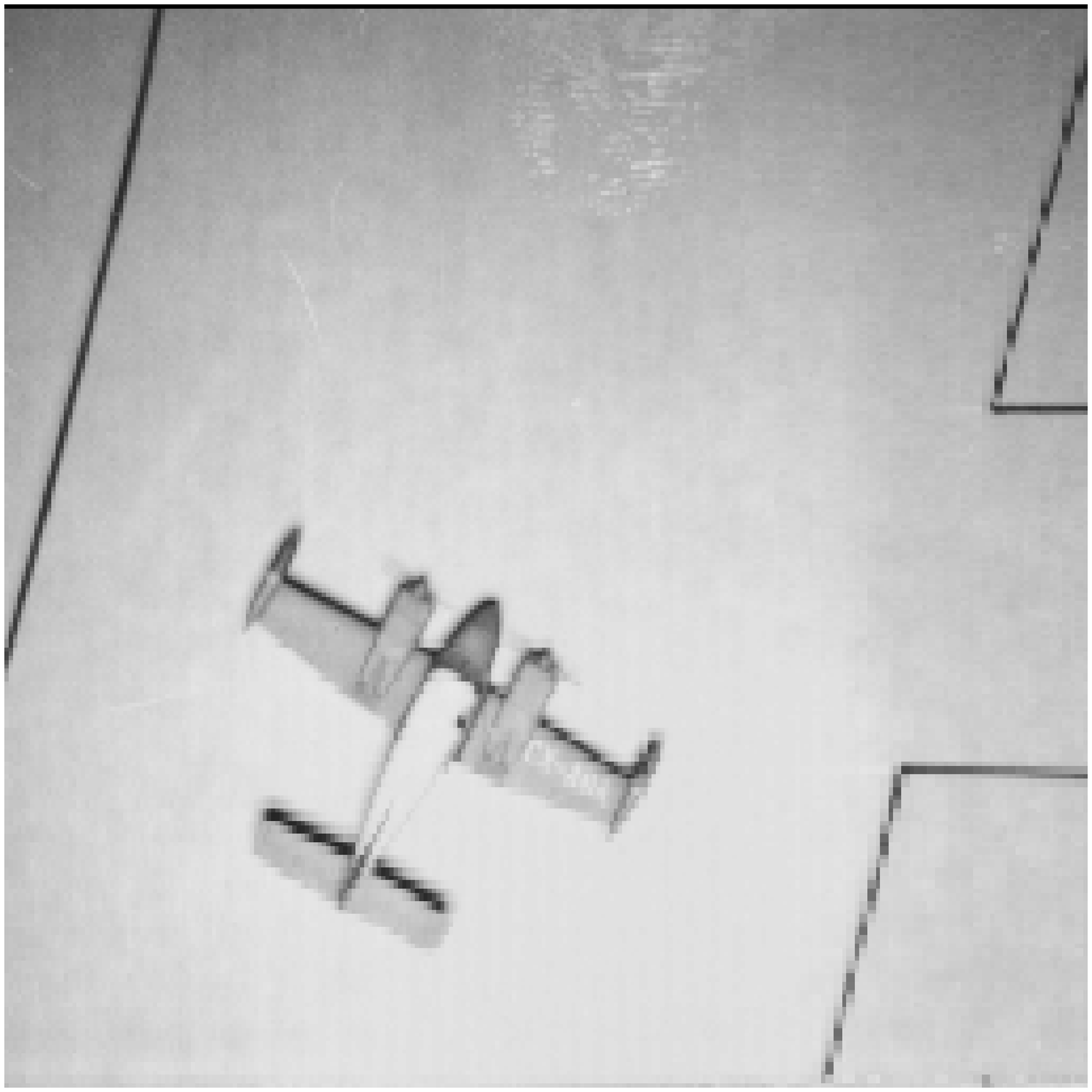,width=.45\linewidth}
\quad
\epsfig{file=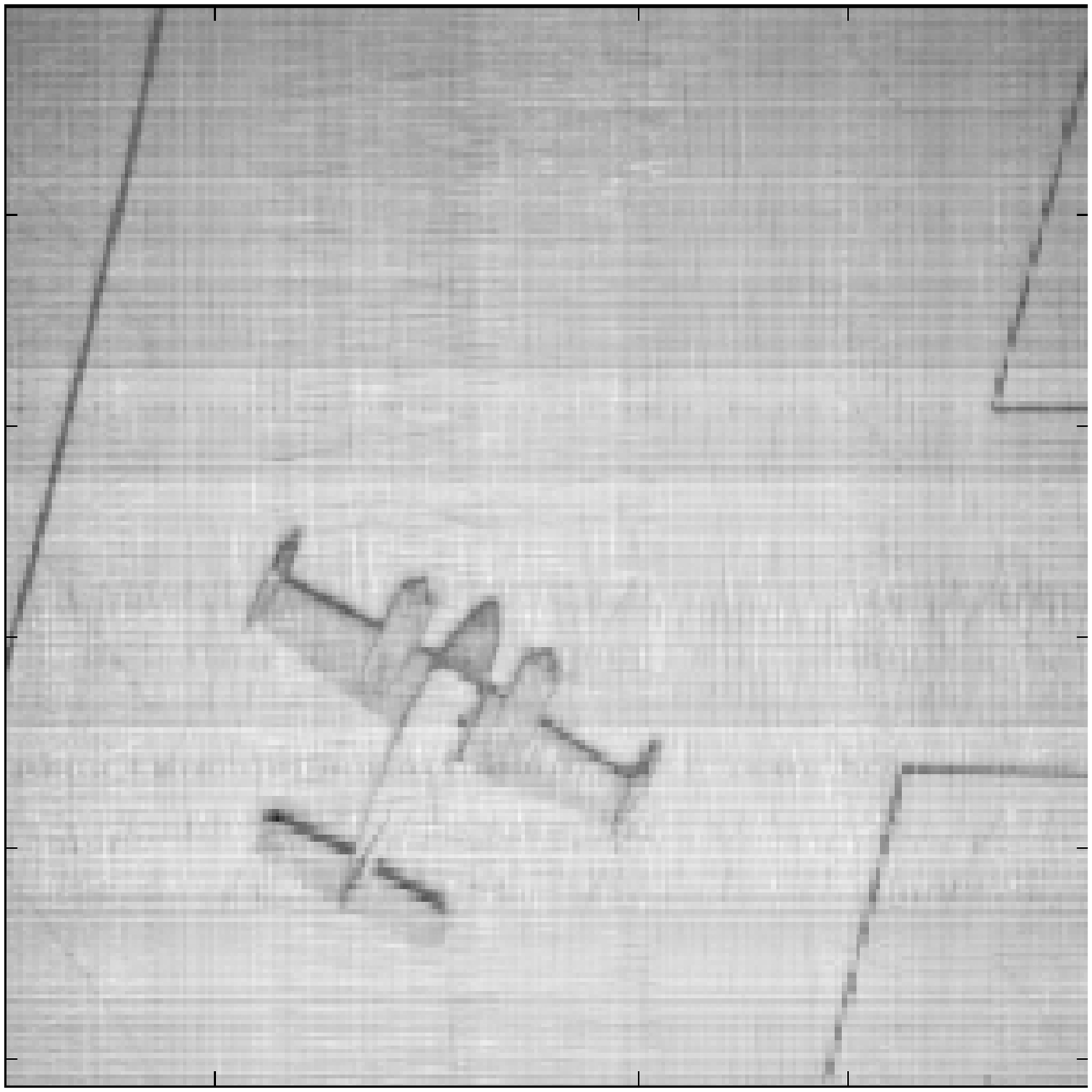,width=.45\linewidth}  }
\subfigure[Aerial]{ \epsfig{file=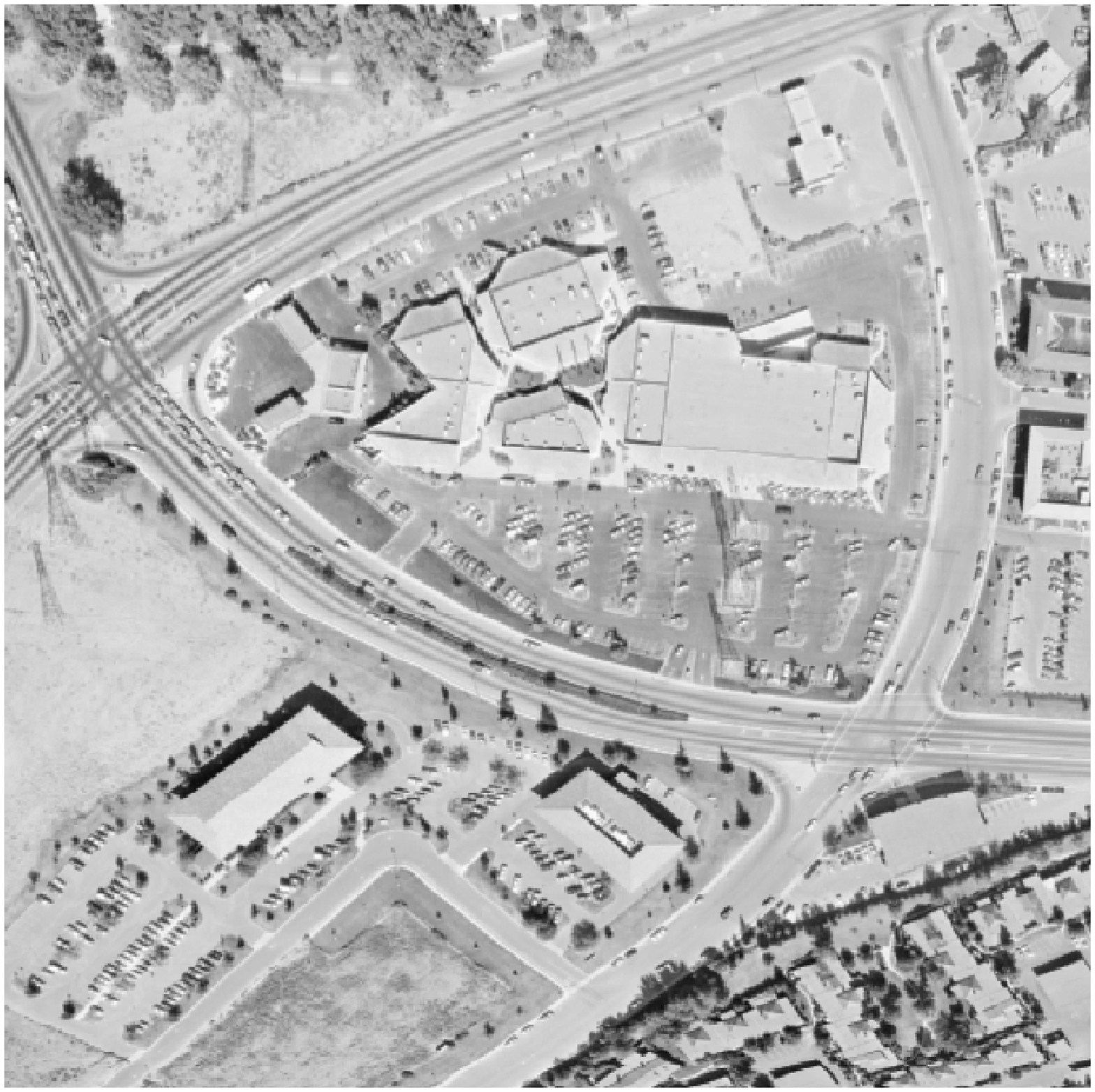,width=.45\linewidth}
\quad
\epsfig{file=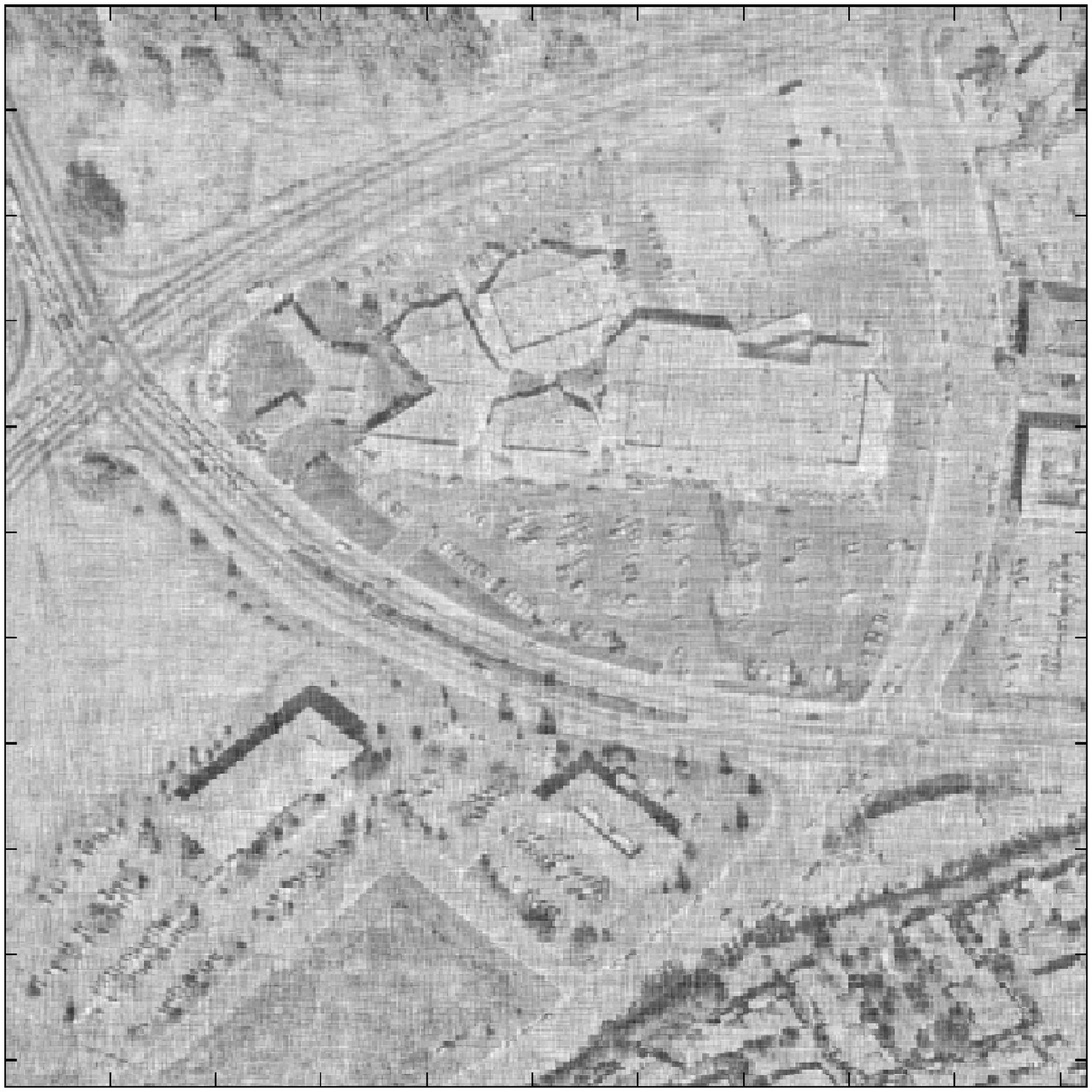,width=.45\linewidth}  }
\caption{Direct and Weak-inverse Transform. The original pictures displayed
on left were direct and weak-inverse transformed via the rounded Hartley transform.
Resulting images are seen on the right.
Since RHT is a quasi-involution, it introduces  noise due to its  nature.}
\label{figset0}
\end{figure}

\begin{figure}
\centering
\subfigure[APC]{
\epsfig{file=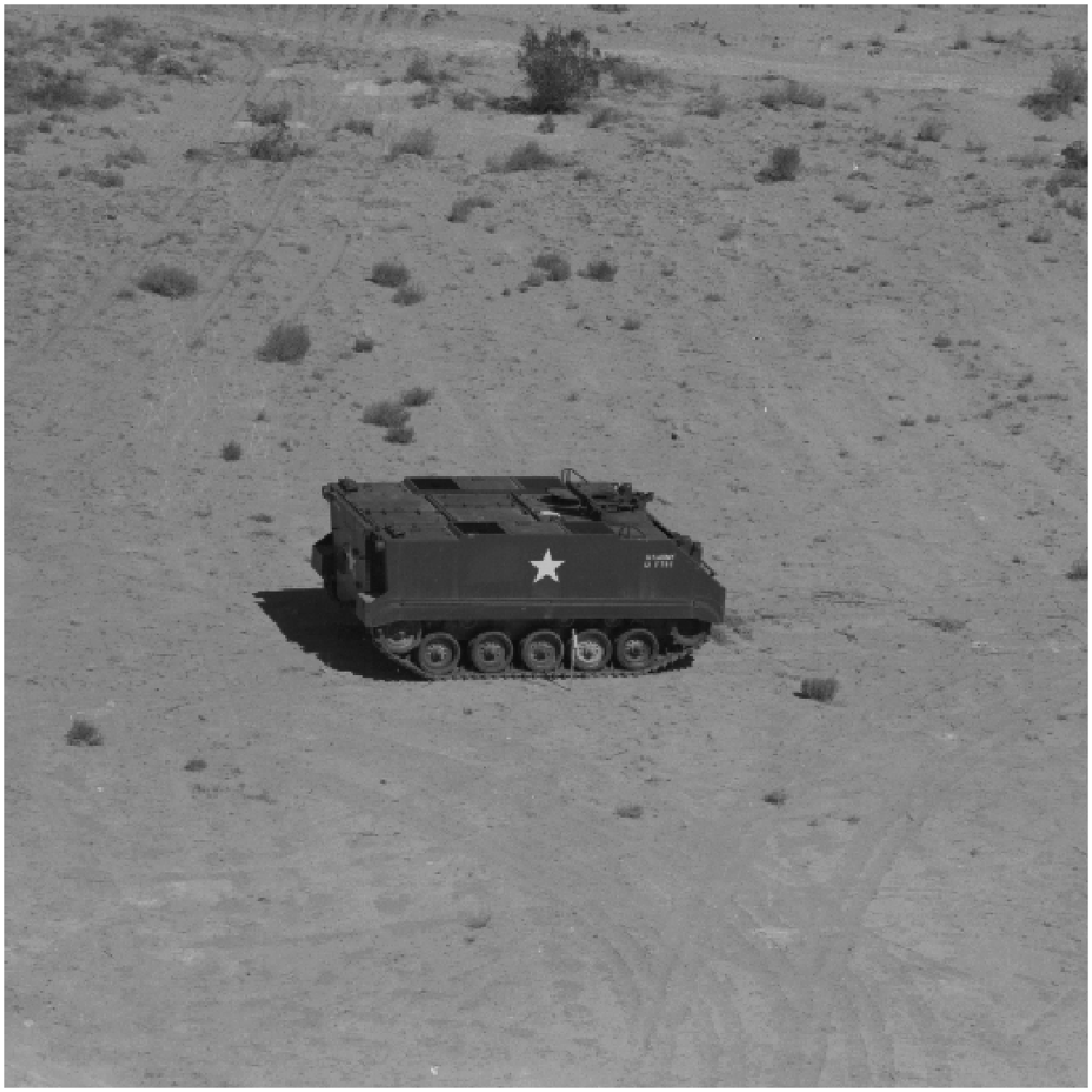,width=.45\linewidth}
\quad
\epsfig{file=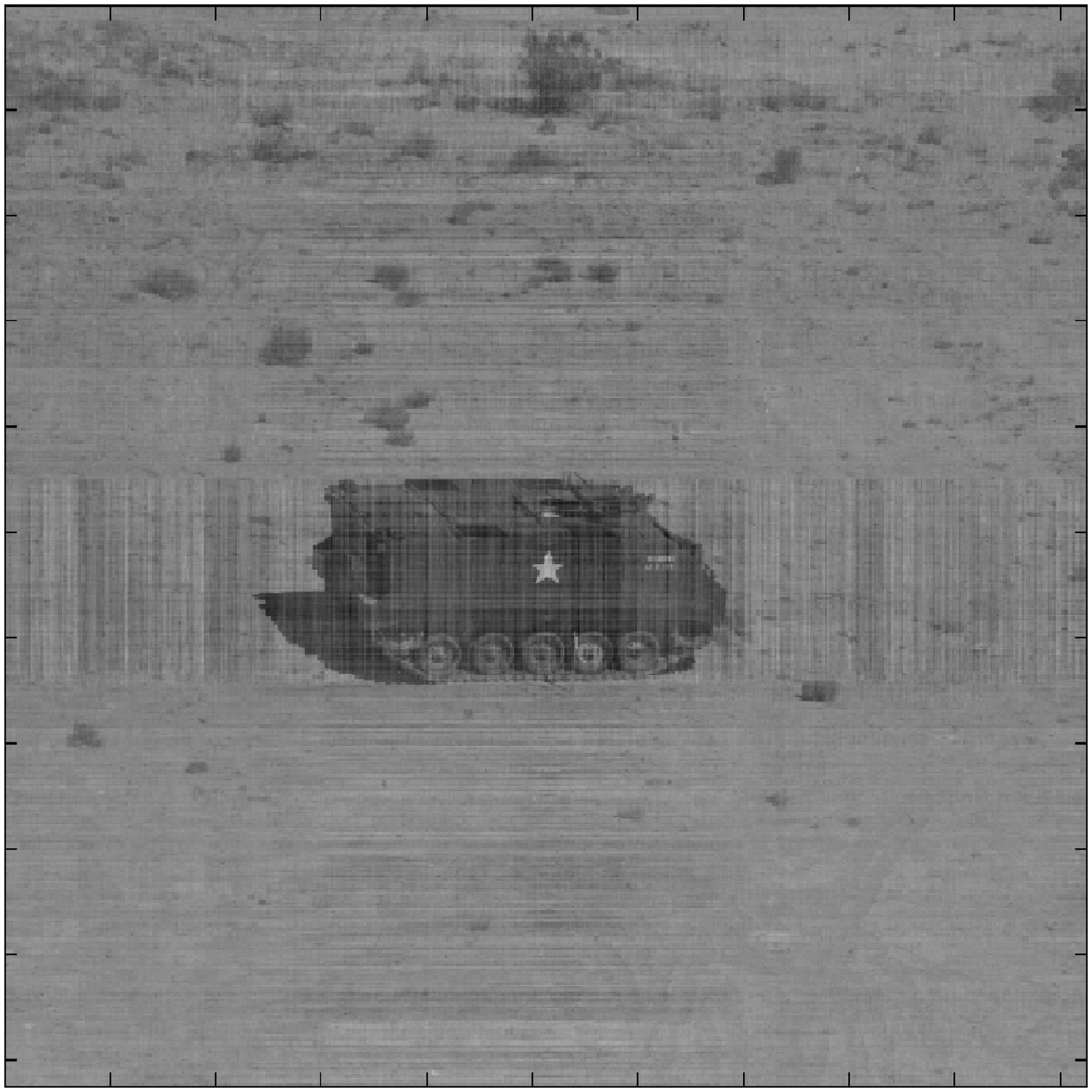,width=.45\linewidth}
}
\subfigure[Tank]{
\epsfig{file=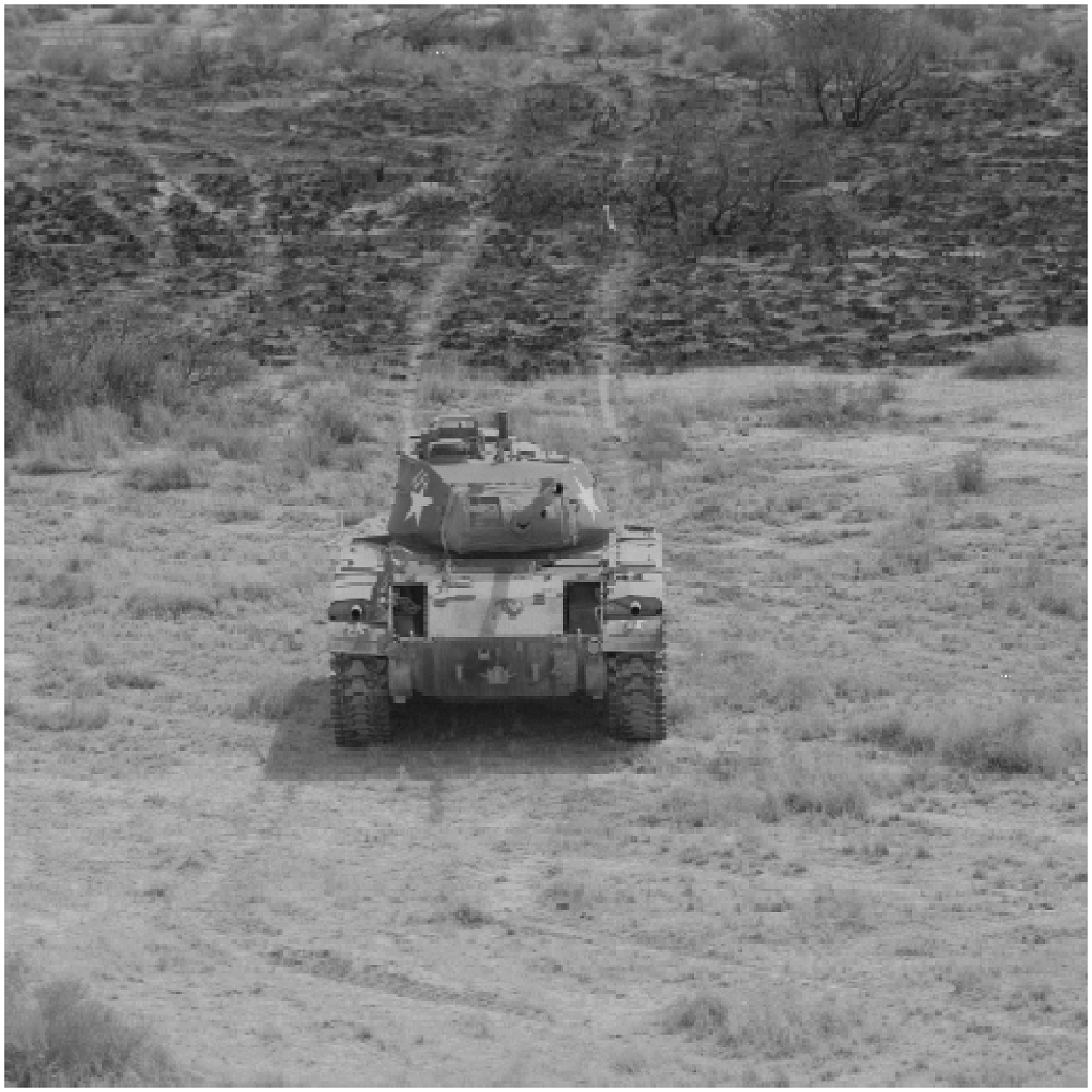,width=.45\linewidth}
\quad
\epsfig{file=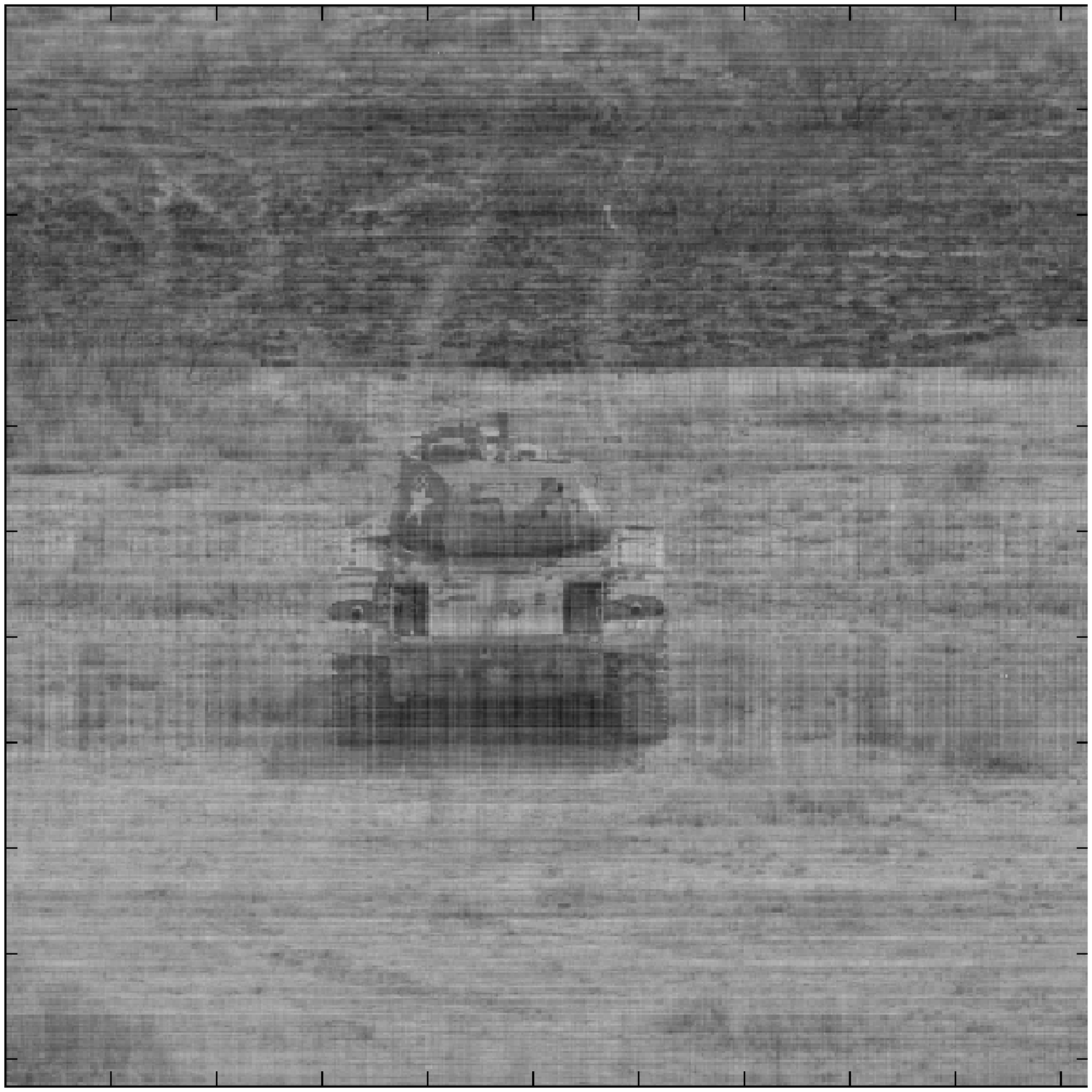,width=.45\linewidth}
}
\caption{Direct and Weak-inverse Transform. Military images.}
\label{figset1}
\end{figure}

Aiming to investigate such degradation
(which follows from the use of weak-inverse), standard images
from Signal and Image Processing Institute Image Database~\cite{uscsipi} at University of
Southern California were analyzed.
Figures~\ref{figset0} and~\ref{figset1} present original images and their respective
recovered images using the weak-inverse transform instead of the (exact) inverse transform.
Program~\ref{program:matlab} lists a naive implementation of 2-D RHT using \mbox{\textsc{Matlab}}.

Table~\ref{tabela:psnr} brings PSNR (Peak Signal-Noise Ratio) of the standard
images after a direct RHT and a weak-inverse RHT.
Observe that the PSNR is image dependent: the quantization noise due to
the rounding depends on the original image characteristics, such as
shape, contrast, dimension etc.

\begin{table}
\caption{PSNR of some standard images after
a direct
and weak-inverse rounded Hartley transform.
All images were obtained from
USC-SIPI Image Database.
In parenthesis, image ID number.}
\label{tabela:psnr}
\begin{minipage}{\linewidth}
\centering
\begin{tabular}{ccc}
\\
\hline
Image & Dimension (pixels) & PSNR (dB) \\
\hline
{Moon surface} (\texttt{5.1.09}) & 256 $\times$ 256 &  26.5522  \\
{Airplane} (\texttt{5.1.11}) & 256 $\times$ 256 & 25.7277 \\
{Aerial} (\texttt{5.2.09} & 512 $\times$ 512 & 22.2006\\
{APC} (\texttt{7.1.08}) & 512 $\times$ 512 & 27.3035\\
{Tank} (\texttt{7.1.09}) & 512 $\times$ 512 & 24.4590\\
\hline
\end{tabular}
\end{minipage}
\end{table}

\section{Connection with Fourier and Walsh/Hadamard Transform}

As final comments on RHT, we present some relationship
between this new transform and other well-known transforms such
as Fourier and Walsh/Hadamard transforms.

Since DHT can be used to compute de DFT~\cite{Bracewell2}, and
the RHT furnishes a estimate for DHT, we can use RHT to
derive a rough --- but fast --- evaluation of Fourier spectrum.

de Oliveira and co-workers~\cite{Oliveira:Multilayer}
found a relationship between
discrete Hartley transform and Hadamard transform.
Such link was exploited to derive new fast algorithms~\cite{Oliveira:Factorization,
Oliveira:Multilayer,
Cintra:Bifuncional}.
In the present framework, we are led to following conjecture:

\begin{conjecture}\label{espirito}
Let $n$ be a power of 2.
The matrix $\hq_{n}$
is identical to Walsh/Hadamard matrix of same order,
except for null elements and for a permutation of columns.
\endproof
\end{conjecture}
For example, the column permutation that, except for zero elements, converts an 8-point rounded
Hartley matrix into a Walsh transformation is $(4\  8)$ (cyclic notation).

\section{Conclusions}

Discrete Hartley transform has long been used in practical applications.
It is real-valued self-inverse transform, more symmetrical
than the DFT~\cite{Bracewell2}.

A new multiplication-free transform derived from DHT is introduced, the RHT, which
kept many properties of discrete Hartley transform.
In spite of not being involutional, it is shown that RHT
exhibits quasi-involutional property, a new concept
derived from the periodicity of matrices.

The quasi-involutional property was induced from weak-inverse definition.
Instead of using the (true) inverse transform, the RHT is viewed as an involutional transform,
allowing the use of direct (multiplication-free) to evaluate the inverse.
Thus, the software/hardware to be used in the computation
of both the direct and the
inverse RHT becomes exactly the same.
The price to be paid by not using the exact inverse transform
is some degradation when recovering original signal.

Fast algorithms to compute RHT are presented showing embedded properties.
The 2-D RHT is also defined, allowing to analyze the
 effects of this approach on standard images.
Despite of SNR loss, RHT can be very interesting for applications involving
image monitoring associated to decision making, such as military applications
or medical imaging.

RHT is offered as an efficient way to compute real-time initial estimations of spectral
evaluations.
Refinement algorithms can be used to improve the image or spectral estimation, when
necessary.
A class of refinement algorithms for this particularly transform is now our object of
investigation.

{\small
\singlespacing
\bibliographystyle{siam}
\bibliography{ref}
}

\end{document}

%% file: qcas_e_as_transformadas.pstex_t
\begin{picture}(0,0)%
\epsfig{file=qcas_e_as_transformadas.pstex}%
\end{picture}%
\setlength{\unitlength}{4144sp}%
\begingroup\makeatletter\ifx\SetFigFont\undefined%
\gdef\SetFigFont#1#2#3#4#5{%
  \reset@font\fontsize{#1}{#2pt}%
  \fontfamily{#3}\fontseries{#4}\fontshape{#5}%
  \selectfont}%
\fi\endgroup%
\begin{picture}(1914,1014)(439,-613)
\put(1081,-511){\makebox(0,0)[lb]{\smash{\SetFigFont{6}{7.2}{\familydefault}{\mddefault}{\updefault}\textsf{Dee-Jeoti}}}}
\put(541,254){\makebox(0,0)[lb]{\smash{\SetFigFont{8}{9.6}{\rmdefault}{\mddefault}{\updefault}Discrete Transforms}}}
\put(1621,-16){\makebox(0,0)[lb]{\smash{\SetFigFont{8}{9.6}{\rmdefault}{\mddefault}{\updefault}Free-multip.}}}
\put(1756,-106){\makebox(0,0)[lb]{\smash{\SetFigFont{6}{7.2}{\rmdefault}{\mddefault}{\updefault}\textsf{Walsh}}}}
\put(1846,209){\makebox(0,0)[lb]{\smash{\SetFigFont{6}{7.2}{\rmdefault}{\mddefault}{\updefault}\textsf{Haar}}}}
\put(541, 29){\makebox(0,0)[lb]{\smash{\SetFigFont{8}{9.6}{\rmdefault}{\mddefault}{\updefault}Trigonometric}}}
\put(586,-106){\makebox(0,0)[lb]{\smash{\SetFigFont{6}{7.2}{\rmdefault}{\mddefault}{\updefault}\textsf{DFT}}}}
\put(721,-286){\makebox(0,0)[lb]{\smash{\SetFigFont{6}{7.2}{\rmdefault}{\mddefault}{\updefault}\textsf{DHT}}}}
\put(1306,-286){\makebox(0,0)[lb]{\smash{\SetFigFont{6}{7.2}{\rmdefault}{\mddefault}{\updefault}\textsf{RHT}}}}
\put(1801,-511){\makebox(0,0)[lb]{\smash{\SetFigFont{8}{9.6}{\rmdefault}{\mddefault}{\updefault}Approx.}}}
\end{picture}

%% file: simple_example.latex
% GNUPLOT: LaTeX picture
\setlength{\unitlength}{0.240900pt}
\ifx\plotpoint\undefined\newsavebox{\plotpoint}\fi
\sbox{\plotpoint}{\rule[-0.200pt]{0.400pt}{0.400pt}}%
\begin{picture}(974,675)(0,0)
\font\gnuplot=cmr10 at 10pt
\gnuplot
\sbox{\plotpoint}{\rule[-0.200pt]{0.400pt}{0.400pt}}%
\put(181.0,123.0){\rule[-0.200pt]{4.818pt}{0.400pt}}
\put(161,123){\makebox(0,0)[r]{-0.5}}
\put(933.0,123.0){\rule[-0.200pt]{4.818pt}{0.400pt}}
\put(181.0,196.0){\rule[-0.200pt]{4.818pt}{0.400pt}}
\put(161,196){\makebox(0,0)[r]{0}}
\put(933.0,196.0){\rule[-0.200pt]{4.818pt}{0.400pt}}
\put(181.0,269.0){\rule[-0.200pt]{4.818pt}{0.400pt}}
\put(161,269){\makebox(0,0)[r]{0.5}}
\put(933.0,269.0){\rule[-0.200pt]{4.818pt}{0.400pt}}
\put(181.0,342.0){\rule[-0.200pt]{4.818pt}{0.400pt}}
\put(161,342){\makebox(0,0)[r]{1}}
\put(933.0,342.0){\rule[-0.200pt]{4.818pt}{0.400pt}}
\put(181.0,416.0){\rule[-0.200pt]{4.818pt}{0.400pt}}
\put(161,416){\makebox(0,0)[r]{1.5}}
\put(933.0,416.0){\rule[-0.200pt]{4.818pt}{0.400pt}}
\put(181.0,489.0){\rule[-0.200pt]{4.818pt}{0.400pt}}
\put(161,489){\makebox(0,0)[r]{2}}
\put(933.0,489.0){\rule[-0.200pt]{4.818pt}{0.400pt}}
\put(181.0,562.0){\rule[-0.200pt]{4.818pt}{0.400pt}}
\put(161,562){\makebox(0,0)[r]{2.5}}
\put(933.0,562.0){\rule[-0.200pt]{4.818pt}{0.400pt}}
\put(181.0,635.0){\rule[-0.200pt]{4.818pt}{0.400pt}}
\put(161,635){\makebox(0,0)[r]{3}}
\put(933.0,635.0){\rule[-0.200pt]{4.818pt}{0.400pt}}
\put(181.0,123.0){\rule[-0.200pt]{0.400pt}{4.818pt}}
\put(181,82){\makebox(0,0){0}}
\put(181.0,615.0){\rule[-0.200pt]{0.400pt}{4.818pt}}
\put(374.0,123.0){\rule[-0.200pt]{0.400pt}{4.818pt}}
\put(374,82){\makebox(0,0){16}}
\put(374.0,615.0){\rule[-0.200pt]{0.400pt}{4.818pt}}
\put(567.0,123.0){\rule[-0.200pt]{0.400pt}{4.818pt}}
\put(567,82){\makebox(0,0){32}}
\put(567.0,615.0){\rule[-0.200pt]{0.400pt}{4.818pt}}
\put(760.0,123.0){\rule[-0.200pt]{0.400pt}{4.818pt}}
\put(760,82){\makebox(0,0){48}}
\put(760.0,615.0){\rule[-0.200pt]{0.400pt}{4.818pt}}
\put(953.0,123.0){\rule[-0.200pt]{0.400pt}{4.818pt}}
\put(953,82){\makebox(0,0){64}}
\put(953.0,615.0){\rule[-0.200pt]{0.400pt}{4.818pt}}
\put(181.0,123.0){\rule[-0.200pt]{185.975pt}{0.400pt}}
\put(953.0,123.0){\rule[-0.200pt]{0.400pt}{123.341pt}}
\put(181.0,635.0){\rule[-0.200pt]{185.975pt}{0.400pt}}
\put(40,379){\makebox(0,0){\shortstack{$V_k$,  \\ $\tilde{V}_k$}}}
\put(567,21){\makebox(0,0){$k$}}
\put(181.0,123.0){\rule[-0.200pt]{0.400pt}{123.341pt}}
\put(181,198){\usebox{\plotpoint}}
\put(253,197.67){\rule{2.891pt}{0.400pt}}
\multiput(253.00,197.17)(6.000,1.000){2}{\rule{1.445pt}{0.400pt}}
\put(181.0,198.0){\rule[-0.200pt]{17.345pt}{0.400pt}}
\put(290,198.67){\rule{2.891pt}{0.400pt}}
\multiput(290.00,198.17)(6.000,1.000){2}{\rule{1.445pt}{0.400pt}}
\put(302,199.67){\rule{2.891pt}{0.400pt}}
\multiput(302.00,199.17)(6.000,1.000){2}{\rule{1.445pt}{0.400pt}}
\put(314,200.67){\rule{2.891pt}{0.400pt}}
\multiput(314.00,200.17)(6.000,1.000){2}{\rule{1.445pt}{0.400pt}}
\put(326,201.67){\rule{2.891pt}{0.400pt}}
\multiput(326.00,201.17)(6.000,1.000){2}{\rule{1.445pt}{0.400pt}}
\multiput(338.00,203.61)(2.472,0.447){3}{\rule{1.700pt}{0.108pt}}
\multiput(338.00,202.17)(8.472,3.000){2}{\rule{0.850pt}{0.400pt}}
\multiput(350.00,206.59)(1.033,0.482){9}{\rule{0.900pt}{0.116pt}}
\multiput(350.00,205.17)(10.132,6.000){2}{\rule{0.450pt}{0.400pt}}
\multiput(362.00,212.58)(0.543,0.492){19}{\rule{0.536pt}{0.118pt}}
\multiput(362.00,211.17)(10.887,11.000){2}{\rule{0.268pt}{0.400pt}}
\multiput(374.58,223.00)(0.492,1.401){21}{\rule{0.119pt}{1.200pt}}
\multiput(373.17,223.00)(12.000,30.509){2}{\rule{0.400pt}{0.600pt}}
\multiput(386.58,256.00)(0.492,7.648){21}{\rule{0.119pt}{6.033pt}}
\multiput(385.17,256.00)(12.000,165.478){2}{\rule{0.400pt}{3.017pt}}
\multiput(398.58,434.00)(0.492,6.571){21}{\rule{0.119pt}{5.200pt}}
\multiput(397.17,434.00)(12.000,142.207){2}{\rule{0.400pt}{2.600pt}}
\multiput(410.58,565.41)(0.492,-6.571){21}{\rule{0.119pt}{5.200pt}}
\multiput(409.17,576.21)(12.000,-142.207){2}{\rule{0.400pt}{2.600pt}}
\multiput(422.58,408.96)(0.492,-7.648){21}{\rule{0.119pt}{6.033pt}}
\multiput(421.17,421.48)(12.000,-165.478){2}{\rule{0.400pt}{3.017pt}}
\multiput(434.58,251.02)(0.492,-1.401){21}{\rule{0.119pt}{1.200pt}}
\multiput(433.17,253.51)(12.000,-30.509){2}{\rule{0.400pt}{0.600pt}}
\multiput(446.00,221.92)(0.543,-0.492){19}{\rule{0.536pt}{0.118pt}}
\multiput(446.00,222.17)(10.887,-11.000){2}{\rule{0.268pt}{0.400pt}}
\multiput(458.00,210.93)(1.378,-0.477){7}{\rule{1.140pt}{0.115pt}}
\multiput(458.00,211.17)(10.634,-5.000){2}{\rule{0.570pt}{0.400pt}}
\multiput(471.00,205.95)(2.472,-0.447){3}{\rule{1.700pt}{0.108pt}}
\multiput(471.00,206.17)(8.472,-3.000){2}{\rule{0.850pt}{0.400pt}}
\put(483,202.17){\rule{2.500pt}{0.400pt}}
\multiput(483.00,203.17)(6.811,-2.000){2}{\rule{1.250pt}{0.400pt}}
\put(495,200.67){\rule{2.891pt}{0.400pt}}
\multiput(495.00,201.17)(6.000,-1.000){2}{\rule{1.445pt}{0.400pt}}
\put(507,199.67){\rule{2.891pt}{0.400pt}}
\multiput(507.00,200.17)(6.000,-1.000){2}{\rule{1.445pt}{0.400pt}}
\put(265.0,199.0){\rule[-0.200pt]{6.022pt}{0.400pt}}
\put(543,198.67){\rule{2.891pt}{0.400pt}}
\multiput(543.00,199.17)(6.000,-1.000){2}{\rule{1.445pt}{0.400pt}}
\put(519.0,200.0){\rule[-0.200pt]{5.782pt}{0.400pt}}
\put(579,198.67){\rule{2.891pt}{0.400pt}}
\multiput(579.00,198.17)(6.000,1.000){2}{\rule{1.445pt}{0.400pt}}
\put(555.0,199.0){\rule[-0.200pt]{5.782pt}{0.400pt}}
\put(615,199.67){\rule{2.891pt}{0.400pt}}
\multiput(615.00,199.17)(6.000,1.000){2}{\rule{1.445pt}{0.400pt}}
\put(627,200.67){\rule{2.891pt}{0.400pt}}
\multiput(627.00,200.17)(6.000,1.000){2}{\rule{1.445pt}{0.400pt}}
\put(639,202.17){\rule{2.500pt}{0.400pt}}
\multiput(639.00,201.17)(6.811,2.000){2}{\rule{1.250pt}{0.400pt}}
\multiput(651.00,204.61)(2.695,0.447){3}{\rule{1.833pt}{0.108pt}}
\multiput(651.00,203.17)(9.195,3.000){2}{\rule{0.917pt}{0.400pt}}
\multiput(664.00,207.59)(1.267,0.477){7}{\rule{1.060pt}{0.115pt}}
\multiput(664.00,206.17)(9.800,5.000){2}{\rule{0.530pt}{0.400pt}}
\multiput(676.00,212.58)(0.543,0.492){19}{\rule{0.536pt}{0.118pt}}
\multiput(676.00,211.17)(10.887,11.000){2}{\rule{0.268pt}{0.400pt}}
\multiput(688.58,223.00)(0.492,1.401){21}{\rule{0.119pt}{1.200pt}}
\multiput(687.17,223.00)(12.000,30.509){2}{\rule{0.400pt}{0.600pt}}
\multiput(700.58,256.00)(0.492,7.648){21}{\rule{0.119pt}{6.033pt}}
\multiput(699.17,256.00)(12.000,165.478){2}{\rule{0.400pt}{3.017pt}}
\multiput(712.58,434.00)(0.492,6.571){21}{\rule{0.119pt}{5.200pt}}
\multiput(711.17,434.00)(12.000,142.207){2}{\rule{0.400pt}{2.600pt}}
\multiput(724.58,565.41)(0.492,-6.571){21}{\rule{0.119pt}{5.200pt}}
\multiput(723.17,576.21)(12.000,-142.207){2}{\rule{0.400pt}{2.600pt}}
\multiput(736.58,408.96)(0.492,-7.648){21}{\rule{0.119pt}{6.033pt}}
\multiput(735.17,421.48)(12.000,-165.478){2}{\rule{0.400pt}{3.017pt}}
\multiput(748.58,251.02)(0.492,-1.401){21}{\rule{0.119pt}{1.200pt}}
\multiput(747.17,253.51)(12.000,-30.509){2}{\rule{0.400pt}{0.600pt}}
\multiput(760.00,221.92)(0.543,-0.492){19}{\rule{0.536pt}{0.118pt}}
\multiput(760.00,222.17)(10.887,-11.000){2}{\rule{0.268pt}{0.400pt}}
\multiput(772.00,210.93)(1.033,-0.482){9}{\rule{0.900pt}{0.116pt}}
\multiput(772.00,211.17)(10.132,-6.000){2}{\rule{0.450pt}{0.400pt}}
\multiput(784.00,204.95)(2.472,-0.447){3}{\rule{1.700pt}{0.108pt}}
\multiput(784.00,205.17)(8.472,-3.000){2}{\rule{0.850pt}{0.400pt}}
\put(796,201.67){\rule{2.891pt}{0.400pt}}
\multiput(796.00,202.17)(6.000,-1.000){2}{\rule{1.445pt}{0.400pt}}
\put(808,200.67){\rule{2.891pt}{0.400pt}}
\multiput(808.00,201.17)(6.000,-1.000){2}{\rule{1.445pt}{0.400pt}}
\put(820,199.67){\rule{2.891pt}{0.400pt}}
\multiput(820.00,200.17)(6.000,-1.000){2}{\rule{1.445pt}{0.400pt}}
\put(832,198.67){\rule{2.891pt}{0.400pt}}
\multiput(832.00,199.17)(6.000,-1.000){2}{\rule{1.445pt}{0.400pt}}
\put(591.0,200.0){\rule[-0.200pt]{5.782pt}{0.400pt}}
\put(869,197.67){\rule{2.891pt}{0.400pt}}
\multiput(869.00,198.17)(6.000,-1.000){2}{\rule{1.445pt}{0.400pt}}
\put(844.0,199.0){\rule[-0.200pt]{6.022pt}{0.400pt}}
\put(881.0,198.0){\rule[-0.200pt]{14.454pt}{0.400pt}}
\put(181,198){\usebox{\plotpoint}}
\multiput(181,198)(10.298,18.021){2}{\usebox{\plotpoint}}
\multiput(193,219)(5.239,-20.083){2}{\usebox{\plotpoint}}
\put(211.27,182.40){\usebox{\plotpoint}}
\put(222.36,182.07){\usebox{\plotpoint}}
\put(236.02,167.49){\usebox{\plotpoint}}
\multiput(241,165)(3.042,20.531){4}{\usebox{\plotpoint}}
\multiput(253,246)(4.754,-20.204){2}{\usebox{\plotpoint}}
\put(270.08,196.95){\usebox{\plotpoint}}
\put(288.60,206.18){\usebox{\plotpoint}}
\put(296.91,189.16){\usebox{\plotpoint}}
\put(305.99,181.32){\usebox{\plotpoint}}
\put(321.30,193.22){\usebox{\plotpoint}}
\put(336.85,205.75){\usebox{\plotpoint}}
\multiput(350,216)(4.503,20.261){3}{\usebox{\plotpoint}}
\multiput(362,270)(5.135,-20.110){2}{\usebox{\plotpoint}}
\multiput(374,223)(5.830,19.920){2}{\usebox{\plotpoint}}
\multiput(386,264)(1.773,20.680){7}{\usebox{\plotpoint}}
\multiput(398,404)(2.154,20.643){6}{\usebox{\plotpoint}}
\multiput(410,519)(2.211,-20.637){5}{\usebox{\plotpoint}}
\multiput(422,407)(2.065,-20.652){6}{\usebox{\plotpoint}}
\multiput(434,287)(3.412,-20.473){3}{\usebox{\plotpoint}}
\multiput(446,215)(9.601,-18.402){2}{\usebox{\plotpoint}}
\put(469.60,203.60){\usebox{\plotpoint}}
\put(480.31,188.70){\usebox{\plotpoint}}
\put(488.24,198.41){\usebox{\plotpoint}}
\put(495.78,216.41){\usebox{\plotpoint}}
\multiput(507,208)(4.503,20.261){3}{\usebox{\plotpoint}}
\multiput(519,262)(5.964,-19.880){2}{\usebox{\plotpoint}}
\put(542.18,214.55){\usebox{\plotpoint}}
\put(557.05,200.66){\usebox{\plotpoint}}
\put(577.52,200.75){\usebox{\plotpoint}}
\put(592.30,214.87){\usebox{\plotpoint}}
\multiput(603,222)(5.964,19.880){2}{\usebox{\plotpoint}}
\multiput(615,262)(4.503,-20.261){3}{\usebox{\plotpoint}}
\put(638.68,216.76){\usebox{\plotpoint}}
\put(645.95,197.87){\usebox{\plotpoint}}
\put(654.15,189.10){\usebox{\plotpoint}}
\put(665.39,203.49){\usebox{\plotpoint}}
\multiput(676,192)(9.601,18.402){2}{\usebox{\plotpoint}}
\multiput(688,215)(3.412,20.473){3}{\usebox{\plotpoint}}
\multiput(700,287)(2.065,20.652){6}{\usebox{\plotpoint}}
\multiput(712,407)(2.211,20.637){5}{\usebox{\plotpoint}}
\multiput(724,519)(2.154,-20.643){6}{\usebox{\plotpoint}}
\multiput(736,404)(1.773,-20.680){7}{\usebox{\plotpoint}}
\multiput(748,264)(5.830,-19.920){2}{\usebox{\plotpoint}}
\multiput(760,223)(5.135,20.110){2}{\usebox{\plotpoint}}
\multiput(772,270)(4.503,-20.261){3}{\usebox{\plotpoint}}
\put(797.66,205.20){\usebox{\plotpoint}}
\put(813.44,193.09){\usebox{\plotpoint}}
\put(828.46,180.72){\usebox{\plotpoint}}
\put(837.36,189.86){\usebox{\plotpoint}}
\put(846.09,205.87){\usebox{\plotpoint}}
\put(864.72,196.78){\usebox{\plotpoint}}
\multiput(869,195)(4.754,20.204){2}{\usebox{\plotpoint}}
\multiput(881,246)(3.042,-20.531){4}{\usebox{\plotpoint}}
\put(898.70,167.85){\usebox{\plotpoint}}
\put(912.06,182.76){\usebox{\plotpoint}}
\put(923.18,181.72){\usebox{\plotpoint}}
\multiput(929,173)(5.239,20.083){2}{\usebox{\plotpoint}}
\put(941,219){\usebox{\plotpoint}}
\end{picture}

%% file: diag_qcas.pstex_t
\begin{picture}(0,0)%
\epsfig{file=diag_qcas.pstex}%
\end{picture}%
\setlength{\unitlength}{4144sp}%
\begingroup\makeatletter\ifx\SetFigFont\undefined%
\gdef\SetFigFont#1#2#3#4#5{%
  \reset@font\fontsize{#1}{#2pt}%
  \fontfamily{#3}\fontseries{#4}\fontshape{#5}%
  \selectfont}%
\fi\endgroup%
\begin{picture}(2340,1383)(271,-1438)
\put(1486,-151){\makebox(0,0)[lb]{\smash{\SetFigFont{9}{10.8}{\familydefault}{\mddefault}{\updefault}${\pmb{\mathsf{H}}}_n$}}}
\put(1396,-1186){\makebox(0,0)[lb]{\smash{\SetFigFont{9}{10.8}{\familydefault}{\mddefault}{\updefault}${\pmb{\mathsf{H}}}_n$}}}
\put(2611,-781){\makebox(0,0)[lb]{\smash{\SetFigFont{9}{10.8}{\familydefault}{\mddefault}{\updefault}${\pmb{\mathsf{V}}}$}}}
\put(586,-511){\makebox(0,0)[lb]{\smash{\SetFigFont{9}{10.8}{\familydefault}{\mddefault}{\updefault}${\mathbf{v}}$}}}
\put(586,-1186){\makebox(0,0)[lb]{\smash{\SetFigFont{9}{10.8}{\familydefault}{\mddefault}{\updefault}$\pmb{{\mathsf{v}}}$}}}
\put(1441,-331){\makebox(0,0)[lb]{\smash{\SetFigFont{8}{9.6}{\familydefault}{\mddefault}{\updefault}easy}}}
\put(271,-1366){\makebox(0,0)[lb]{\smash{\SetFigFont{8}{9.6}{\rmdefault}{\mddefault}{\updefault}(``almost'' ${\mathbf{v}}$)}}}
\put(1351,-1411){\makebox(0,0)[lb]{\smash{\SetFigFont{8}{9.6}{\familydefault}{\mddefault}{\updefault}easy}}}
\put(1351,-691){\makebox(0,0)[lb]{\smash{\SetFigFont{9}{10.8}{\familydefault}{\mddefault}{\updefault}${\pmb{\mathsf{H}}}^{-1}_n$}}}
\put(1261,-916){\makebox(0,0)[lb]{\smash{\SetFigFont{8}{9.6}{\familydefault}{\mddefault}{\updefault}harder}}}
\end{picture}

%% file: qdht-16.pstex_t
\begin{picture}(0,0)%
\epsfig{file=qdht-16.pstex}%
\end{picture}%
\setlength{\unitlength}{2859sp}%
\begingroup\makeatletter\ifx\SetFigFont\undefined%
\gdef\SetFigFont#1#2#3#4#5{%
  \reset@font\fontsize{#1}{#2pt}%
  \fontfamily{#3}\fontseries{#4}\fontshape{#5}%
  \selectfont}%
\fi\endgroup%
\begin{picture}(4555,4476)(28,-3660)
\put( 28,-1017){\makebox(0,0)[lb]{\smash{\SetFigFont{6}{7.2}{\familydefault}{\mddefault}{\updefault}$v_6$}}}
\put( 28,683){\makebox(0,0)[lb]{\smash{\SetFigFont{6}{7.2}{\familydefault}{\mddefault}{\updefault}$v_0$}}}
\put( 28,-731){\makebox(0,0)[lb]{\smash{\SetFigFont{6}{7.2}{\familydefault}{\mddefault}{\updefault}$v_5$}}}
\put( 28,-1303){\makebox(0,0)[lb]{\smash{\SetFigFont{6}{7.2}{\familydefault}{\mddefault}{\updefault}$v_7$}}}
\put( 28,412){\makebox(0,0)[lb]{\smash{\SetFigFont{6}{7.2}{\familydefault}{\mddefault}{\updefault}$v_1$}}}
\put( 28,126){\makebox(0,0)[lb]{\smash{\SetFigFont{6}{7.2}{\familydefault}{\mddefault}{\updefault}$v_2$}}}
\put( 28,-160){\makebox(0,0)[lb]{\smash{\SetFigFont{6}{7.2}{\familydefault}{\mddefault}{\updefault}$v_3$}}}
\put( 28,-445){\makebox(0,0)[lb]{\smash{\SetFigFont{6}{7.2}{\familydefault}{\mddefault}{\updefault}$v_4$}}}
\put( 28,-1603){\makebox(0,0)[lb]{\smash{\SetFigFont{6}{7.2}{\familydefault}{\mddefault}{\updefault}$v_8$}}}
\put( 28,-3017){\makebox(0,0)[lb]{\smash{\SetFigFont{6}{7.2}{\familydefault}{\mddefault}{\updefault}$v_{13}$}}}
\put( 28,-3303){\makebox(0,0)[lb]{\smash{\SetFigFont{6}{7.2}{\familydefault}{\mddefault}{\updefault}$v_{14}$}}}
\put( 28,-3589){\makebox(0,0)[lb]{\smash{\SetFigFont{6}{7.2}{\familydefault}{\mddefault}{\updefault}$v_{15}$}}}
\put( 28,-1874){\makebox(0,0)[lb]{\smash{\SetFigFont{6}{7.2}{\familydefault}{\mddefault}{\updefault}$v_9$}}}
\put( 28,-2160){\makebox(0,0)[lb]{\smash{\SetFigFont{6}{7.2}{\familydefault}{\mddefault}{\updefault}$v_{10}$}}}
\put( 28,-2446){\makebox(0,0)[lb]{\smash{\SetFigFont{6}{7.2}{\familydefault}{\mddefault}{\updefault}$v_{11}$}}}
\put( 28,-2731){\makebox(0,0)[lb]{\smash{\SetFigFont{6}{7.2}{\familydefault}{\mddefault}{\updefault}$v_{12}$}}}
\put(4571,698){\makebox(0,0)[lb]{\smash{\SetFigFont{6}{7.2}{\familydefault}{\mddefault}{\updefault}$V_0$}}}
\put(4571,-160){\makebox(0,0)[lb]{\smash{\SetFigFont{6}{7.2}{\familydefault}{\mddefault}{\updefault}$V_{12}$}}}
\put(4571,-445){\makebox(0,0)[lb]{\smash{\SetFigFont{6}{7.2}{\familydefault}{\mddefault}{\updefault}$V_2$}}}
\put(4571,-731){\makebox(0,0)[lb]{\smash{\SetFigFont{6}{7.2}{\familydefault}{\mddefault}{\updefault}$V_{10}$}}}
\put(4571,-1017){\makebox(0,0)[lb]{\smash{\SetFigFont{6}{7.2}{\familydefault}{\mddefault}{\updefault}$V_6$}}}
\put(4571,-1303){\makebox(0,0)[lb]{\smash{\SetFigFont{6}{7.2}{\familydefault}{\mddefault}{\updefault}$V_{14}$}}}
\put(4571,-1588){\makebox(0,0)[lb]{\smash{\SetFigFont{6}{7.2}{\familydefault}{\mddefault}{\updefault}$V_1$}}}
\put(4571,-1874){\makebox(0,0)[lb]{\smash{\SetFigFont{6}{7.2}{\familydefault}{\mddefault}{\updefault}$V_9$}}}
\put(4571,-2160){\makebox(0,0)[lb]{\smash{\SetFigFont{6}{7.2}{\familydefault}{\mddefault}{\updefault}$V_5$}}}
\put(4571,-2446){\makebox(0,0)[lb]{\smash{\SetFigFont{6}{7.2}{\familydefault}{\mddefault}{\updefault}$V_{13}$}}}
\put(4571,-2731){\makebox(0,0)[lb]{\smash{\SetFigFont{6}{7.2}{\familydefault}{\mddefault}{\updefault}$V_3$}}}
\put(4571,412){\makebox(0,0)[lb]{\smash{\SetFigFont{6}{7.2}{\familydefault}{\mddefault}{\updefault}$V_8$}}}
\put(4571,126){\makebox(0,0)[lb]{\smash{\SetFigFont{6}{7.2}{\familydefault}{\mddefault}{\updefault}$V_4$}}}
\put(4571,-3303){\makebox(0,0)[lb]{\smash{\SetFigFont{6}{7.2}{\familydefault}{\mddefault}{\updefault}$V_{15}$}}}
\put(4571,-3589){\makebox(0,0)[lb]{\smash{\SetFigFont{6}{7.2}{\familydefault}{\mddefault}{\updefault}$V_{11}$}}}
\put(4571,-3017){\makebox(0,0)[lb]{\smash{\SetFigFont{6}{7.2}{\familydefault}{\mddefault}{\updefault}$V_7$}}}
\end{picture}